%% file: main.tex
\documentclass[10pt,twocolumn,letterpaper]{article}

\usepackage{cvpr}

\usepackage{times}
\usepackage[T1]{fontenc}
\usepackage[scaled=0.90]{inconsolata}

\usepackage{epsfig}
\usepackage{graphicx}
\usepackage{amsmath}
\usepackage{amssymb}
\usepackage[version=3]{mhchem}
\usepackage{gensymb}
\usepackage{array}
\usepackage{glossaries}
\usepackage{comment}

\usepackage{xcolor}
\definecolor{Belize}{RGB}{41,128,185}
\definecolor{JalapenoRed}{RGB}{183,21,64}
\definecolor{Orchid}{RGB}{176,103,161}
\definecolor{Violet}{RGB}{95,75,139}
\usepackage{cite}
\usepackage[pagebackref=true,breaklinks=true,letterpaper=true,colorlinks,bookmarks=false,allcolors=JalapenoRed]{hyperref}

\makeglossaries
\input{glossary}

\cvprfinalcopy 

\def\cvprPaperID{****} 

\newcommand{\lz}[1]{\textcolor{blue}{LZ-comment: #1}}

\newcommand{\dataset}[1]{OC20}

\ifcvprfinal\fi

\begin{document}

\title{An Introduction to Electrocatalyst Design using Machine Learning \\
for Renewable Energy Storage}

\author{C. Lawrence Zitnick$^{*1}$, Lowik Chanussot$^1$, Abhishek Das$^1$, Siddharth Goyal$^1$, Javier Heras-Domingo$^2$, \\ Caleb Ho$^1$, Weihua Hu$^5$, Thibaut Lavril$^1$, Aini Palizhati$^2$, Morgane Rivi\`{e}re$^1$, Muhammed Shuaibi$^2$, \\ Anuroop Sriram$^1$, Kevin Tran$^2$, Brandon Wood$^4$, Junwoong Yoon$^2$, Devi Parikh$^{1,3}$, Zachary Ulissi$^{*2}$ \\
{\tt\small $^1$Facebook AI Research, $^2$CMU, $^3$Georgia Tech, $^4$NERSC, $^5$Stanford} \\
{\tt\small $^*$corresponding authors}
}

\maketitle

\begin{abstract}
Scalable and cost-effective solutions to renewable energy storage are essential to addressing the world's rising energy needs while reducing climate change. As we increase our reliance on renewable energy sources such as wind and solar, which produce intermittent power, storage is needed to transfer power from times of peak generation to peak demand. This may require the storage of power for hours, days, or months. One solution that offers the potential of scaling to nation-sized grids is the conversion of renewable energy to other fuels, such as hydrogen or methane. To be widely adopted, this process requires cost-effective solutions to running electrochemical reactions. An open challenge is finding low-cost electrocatalysts to drive these reactions at high rates. Through the use of quantum mechanical simulations (density functional theory), new catalyst structures can be tested and evaluated. Unfortunately, the high computational cost of these simulations limits the number of structures that may be tested. The use of machine learning may provide a method to efficiently approximate these calculations, leading to new approaches in finding effective electrocatalysts. In this paper, we provide an introduction to the challenges in finding suitable electrocatalysts, how machine learning may be applied to the problem, and the use of the Open Catalyst Project \dataset{} dataset for model training. 
\end{abstract}

\input{sections/intro}

\input{sections/storage}
\input{sections/electrocatalyst}

\input{sections/ml}

\input{sections/discussion}

{\small
\bibliographystyle{ieeetr}
\bibliography{main}
}
\printglossaries
\end{document}

%% file: glossary.tex
\newglossaryentry{adsorbate}
{
    name=adsorbates,
    description={Reactants or intermediate molecules that are adsorbed by a catalyst during a reaction}
}

\newglossaryentry{electrocatalyst}
{
    name=electrocatalyst,
    description={A catalyst used for electrochemical reactions. Typically, both an anode and a cathode electrocatalyst are used}
}

\newglossaryentry{catalyst}
{
    name=catalyst,
    description={A substance used to increase the rate of a chemical reaction that is not consumed in the process}
}

\newglossaryentry{electrochemical_reaction}
{
    name=electrochemical reaction,
    description={A chemical reaction that generates an electric current}
}

\newglossaryentry{intermediate}
{
    name=intermediate,
    description={A molecule that is the by-product of the reaction's reactants or other intermediates that forms before producing the final product}
}

\newglossaryentry{GFE}
{
    name=Gibbs free energy,
    description={The thermodynamic energy associated with a system that can be used to do work at a constant temperature and pressure}
}

\newglossaryentry{PSH}
{
    name=Pumped-Storage Hydroelectric (PSH),
    description={The energy storage process for pumping water uphill that is later used to generate power through a hydroelectric plant}
}

\newglossaryentry{HES}
{
    name=Hydrogen Energy Storage (HES),
    description={The process for storing energy in the form of Hydrogen}
}

\newglossaryentry{DFT}
{
    name=Density Functional Theory (DFT),
    description={A method for computing the electronic energy of a system of atoms from first principles}
}

\newglossaryentry{fuel_cell}
{
    name=fuel cells,
    description={A device used to convert hydrogen and oxygen into electricity and water}
}

\newglossaryentry{free_energy}
{
    name=reaction's free energy,
    description={The energy released or adsorbed during a chemical reaction}
}

\newglossaryentry{exergonic}
{
    name=exergonic,
    description={A spontaneous reaction with negative free energy}
}

\newglossaryentry{endergonic}
{
    name=endergonic,
    description={A reaction with positive free energy that requires energy to activate}
}

\newglossaryentry{exothermic}
{
    name=exothermic,
    description={A reaction that releases heat}
}

\newglossaryentry{endothermic}
{
    name=endothermic,
    description={A reaction that adsorbs heat}
}

\newglossaryentry{activation_energy}
{
    name=activation energy,
    description={The amount of energy needed to initiate a chemical reaction}
}

\newglossaryentry{adsorption_energy}
{
    name=adsorption energy,
    description={The change in energy of the system before and after the adsorbate bonds with a catalyst}
}

\newglossaryentry{enthalpy}
{
    name=enthalpy,
    description={The energy contained in the bonds between atoms}
}

\newglossaryentry{relaxed_system}
{
    name=relaxed system,
    description={A system whose atom positions result in a local minimum in its energy}
}

\newglossaryentry{electronic_energy}
{
    name=electronic energy,
    description={The energy of a system of atoms computed by DFT. This includes the potential and kinetic energies, as well as, the quantum mechanical exchange energy and total correlation energy of the electrons}
}

\newglossaryentry{slab}
{
    name=slab,
    description={A set of atoms corresponding to a single crystalline structure of a catalyst. Multiple slabs may be tiled to create a catalyst surface}
}

%% file: sections/intro.tex
\section{Introduction}
The increased use of renewable energy is essential for addressing the world's rising energy needs while reducing climate change. Currently, renewable energy such as solar, wind and hydro generates $17\%$ of the US's and $28\%$ of the world's electricity \cite{EPM-US, IEO-Int}. While hydro accounts for $62\%$ of this generation, it is limited in its ability to expand further due to resource availability and environmental concerns. Meanwhile, solar and wind are projected by 2050 to expand significantly and account for over $70\%$ of renewable energy generation \cite{IEO-Int}. With this expansion, a unique and potentially limiting challenge arises. Unlike traditional power sources, such as coal, natural gas, and nuclear, power provided by solar and wind is intermittent.

\begin{figure}
	\begin{center}
		\includegraphics[width=0.98\columnwidth]{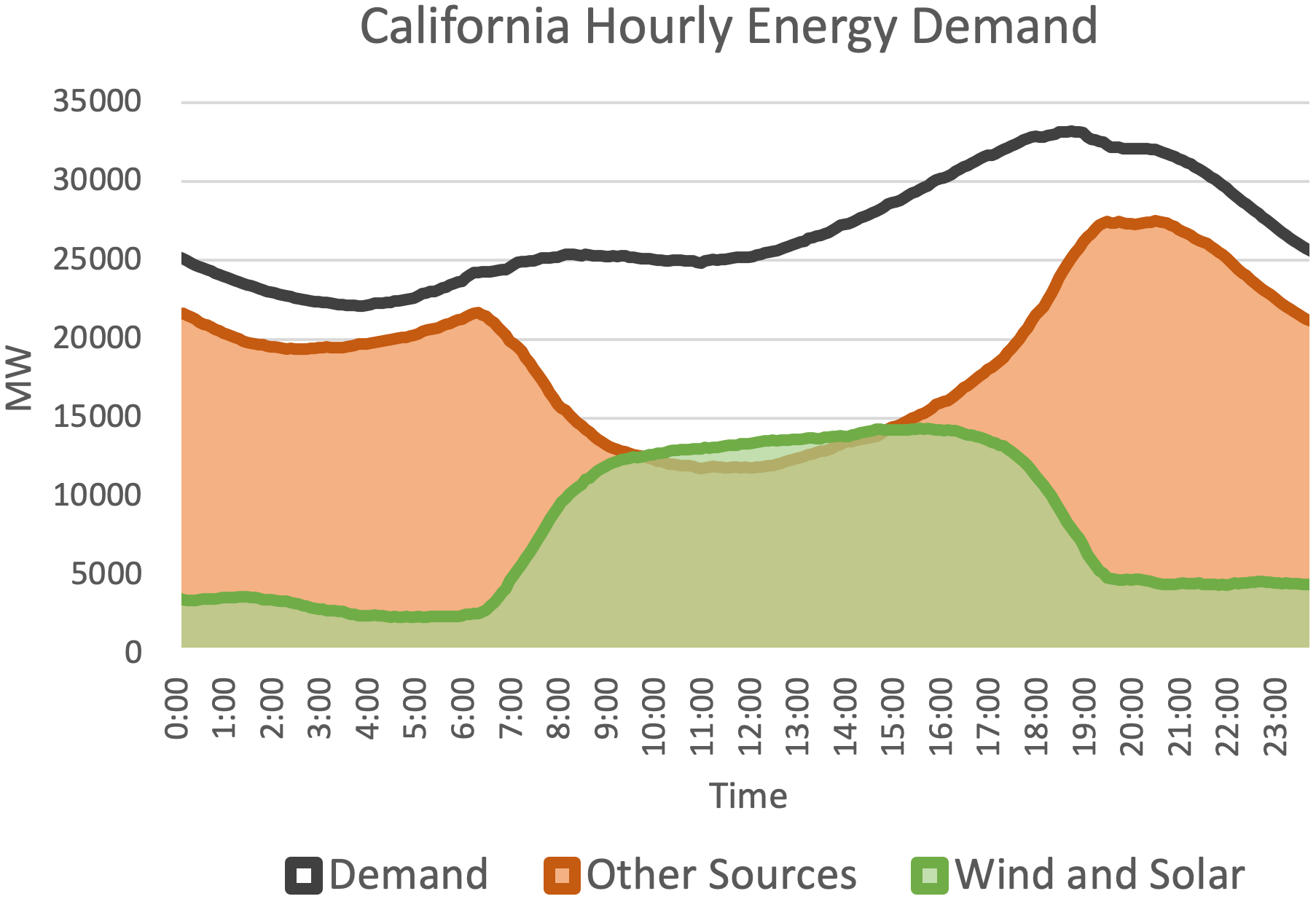}
	\end{center}
	\caption{Hourly electricity demand for California on a typical summer day (August 6th, 2020) as reported by California ISO \cite{CAISO}. The demand (black line) peeks around 19:00, the same time that output from solar and wind power (green line) declines. The remaining demand after the use of wind and solar power is shown by the orange line. Note that energy storage is needed if solar and wind are to meet demand without excess energy loss.}
	\label{fig:hourly}
\end{figure}

The wind does not always blow, and the sun sets. During these periods, wind and solar plants do not generate any power. Notably, energy demand typically peaks in the early mornings (8am) and evenings (6-7pm) well outside the peak generating time (noon) for solar power \cite{power_monitor}, Figure \ref{fig:hourly}. On the other hand, during peak mid-day solar power generation, the power from other more conventional sources must be reduced to balance supply and demand. Adjustments in output from conventional plants may not always be able to match the variability of solar and wind power generation (\eg due to the unpredictability of rapid weather fluctuations), resulting in an excess of supply and wasted energy \cite{cochran2014flexibility}. Without further advances, it is estimated the overall penetration of solar power is capped at $30\%$ and costs start to rise after $20\%$ penetration \cite{Sivaram19}. Currently, areas such as California that generate $14\%$ to $19\%$ of their power from solar \cite{CA_solar_CEC,CA_solar_SEIA} are approaching these levels.

The challenges raised by intermittent power can be addressed through several means, including demand response \cite{Load_Flex, DR}, improved power transmission \cite{Transmission_Madrigal}, and energy storage\cite{amrouche2016overview, auer2012state, 2018iv, Renewable_Storage, ziegler2019storage}. Demand response provides electricity consumers with the ability to shift power demand to time periods with high supply. The transmission of power across large distances can help both transmit power from renewable energy sites that are far away from cities, and smooth fluctuations in the generation of intermittent power by combining multiple sites. While both of these techniques are part of the solution, efficient storage of renewable energy is essential to unlocking the full potential of intermittent renewable energy.

Numerous storage techniques exist, including pumping water uphill (Pumped-Storage Hydroelectricity, PSH) \cite{Hydropower}, batteries \cite{amrouche2016overview}, and \gls{HES} \cite{zhang2016survey,pickard2012addressing,Lohner2013}. HES combines electricity from renewable sources with water to generate hydrogen through a process called electrolysis. The hydrogen is stored and used when needed to generate electricity through \gls{fuel_cell}. As described in greater detail in Section \ref{sec:storage}, each of these storage approaches has trade offs and the appropriate choice is based on the application. PSH and batteries both have high efficiency, but do not scale to the energy storage needs of large grids powered by intermittent renewable energy due to land restrictions (PSH) or cost (batteries). Hydrogen energy storage offers a unique combination of scalability, long-term storage and portability. Specifically, HES can store vast quantities of energy cheaply in the form of hydrogen using the existing gas storage infrastructure. However, other costs associated with HES has limited the adoption of this otherwise very promising technology.

Hydrogen energy storage's high costs result from its relatively low efficiency and the capital costs associated with systems for power conversion (electricity to hydrogen and back to electricity) \cite{steward2009lifecycle,zakeri2015electrical}. The efficiency of HES is estimated to be $35\%$ for round-trip AC to AC \cite{zakeri2015electrical}, and unlikely to exceed $50\%$. Compared to other systems such as PSH, which can achieve $70-80\%$ efficiency, the cost of the electricity used to feed the system is nearly double. Thus the relative overall cost of HES compared to other systems is highly dependent on fluctuations in electricity pricing. Assuming the cost of electricity from renewable energy sources continues to decline (solar costs have reduced 100-fold since 1976 \cite{owidrenewableenergy}) and as the economic necessity of carbon neutral energy sources increases, the attractiveness of HES use should only improve.

\begin{figure*}
	\begin{center}
		\includegraphics[width=0.92\linewidth]{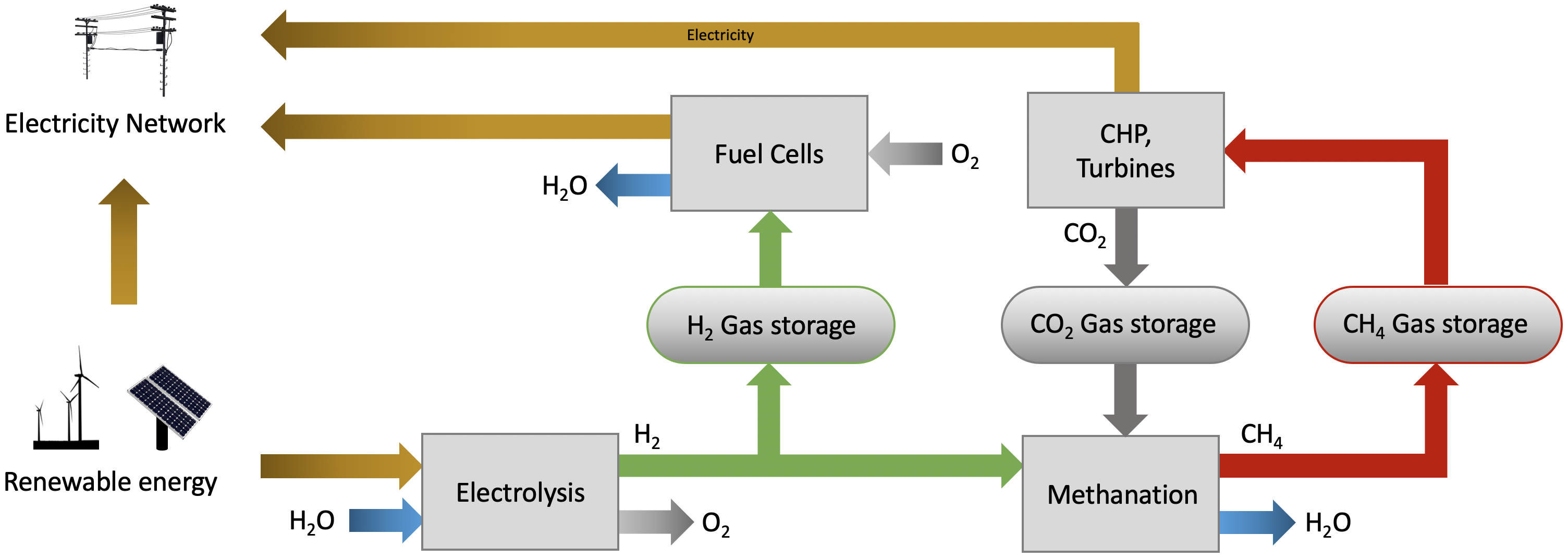}
	\end{center}
	\caption{Diagram of the hydrogen (\ce{H2}) and methane (\ce{CH4}) energy storage process using renewable energy. The stored hydrogen and methane is used to generate electricity when renewable energy generation is not able to meet demand. Renewable energy (bottom left) may either be fed directly into the grid or used to generate hydrogen by electrolysis. The hydrogen is stored for future electricity generation by fuel cells or used in the methanation process to generate methane. The generation of electricity from stored methane can potentially be carbon neutral by recycling the \ce{CO2} for later use in methanation.}
	\label{fig:powercycle}
\end{figure*}

The power conversion systems for HES involve splitting water into its constituent parts (hydrogen and oxygen) using an electrolyzer and converting the hydrogen back to electricity using fuel cells or other methods such as gas-fired turbines and engines (Figure~\ref{fig:powercycle}). The electrolyzer and the fuel cell perform inverse reactions, \ce{2H2O -> 2H2 + O2} and \ce{2H2 + O2 -> 2H2O} respectively. Both of these systems rely on materials called \gls{electrocatalyst}s to increase the rate and efficiency of their respective electrochemical reactions. Unfortunately, common state-of-the-art electrocatalysts use expensive noble metals such as iridium and platinum \cite{sapountzi2017electrocatalysts}. A major challenge is finding low-cost, efficient, and durable electrocatalysts for use in these reactions. If such catalysts can be found, earlier adoption of HES systems could be achieved due to significant cost reductions.

Modern design of \gls{catalyst}s uses computational simulation to determine if a material is suitable for further exploration. The simulation models the interaction of a molecule with an catalyst's surface in a process called adsorption. The adsorbed molecules---i.e., \gls{adsorbate}---are typically the key chemical species involved in the chemical reactions of interest, such as \ce{OH}, \ce{O2}, or \ce{H2O}. The simulation is performed using \gls{DFT}\cite{sholl2011density, parr1980density}. While the foundations of DFT were first introduced in the 1960s \cite{hohenberg1964inhomogeneous,kohn1965self}, it wasn't until the 1990s that advances resulted in practical predictions for quantum chemistry \cite{becke2014perspective}. Now DFT is commonly used to compute the interactions of atoms using a process colloquially called ``relaxation''. In relaxations, DFT combines the locations of the atomic nuclei with quantum mechanics to predict the energy of the system and the forces acting on each atom. The locations of the atomic nuclei are then updated to minimize the energy, consequently changing the electronic distributions and energies. This iterative process continues until the energy of the system reaches a local minima. By examining the energy of the resulting \gls{relaxed_system} (system with lowest energy), we can gain insight into the overall efficiency of the reaction, i.e., how much energy results from or is needed to drive the reaction. Unfortunately, the DFT computation is extremely expensive and scales $O(n^3)$ with the number of electrons in the system, which can range from O(100) to over O(5,000) electrons in practice. The result is computations that take hours or days per relaxation on O(10--100) core CPUs. Beyond DFT, alternative approaches to modeling quantum mechanics include the use of Coupled Cluster Single Double Triple (CCSDT)\cite{noga1987full} that improves on the accuracy of DFT but is rarely used in practice due to its even higher computational cost, and force fields which are computationally very efficient but fall short in accuracy \cite{van2001reaxff,cornell1995second}.

To enable large-scale exploration of new catalysts, efficient computational approximations to DFT are needed. One particularly promising approach is to use machine learning \cite{brockherde2017bypassing}. Given a training dataset of DFT calculations, machine learning techniques, such as Deep Learning (DL), have shown the ability to learn good approximations for certain problems. These techniques take as input the state of the system, including the atom positions, element types, bond information, etc., and predict system properties such as their energy \cite{gilmer2017neural,schutt2017schnet,jorgensen2018neural,xie2018crystal}. Popular datasets for this task include QM9 \cite{ramakrishnan2014quantum} containing over 134k small organic molecules, the Material Project containing 70k inorganic compounds \cite{jain2013commentary}, and ISO17 containing motion trajectories for 129 molecules from QM9 \cite{schutt2017schnet}. Unfortunately, similar datasets relevant to our problem of interest remain limited\cite{tran2018active,tran2019methods}. 

An ideal dataset containing catalysts for renewable energy storage would have two main properties. First, it would be large scale, containing DFT simulations of relevant catalysts and adsorbates.  Second, the dataset would enable the training of models that take as input rough initial starting positions of the system's atoms from which the relaxed geometries and energies could be computed. The large majority of DFT calculations are spent in the iterative computation of relaxed geometries, so efficient approximations of this process are needed. This paper provides an introduction to how the Open Catalyst Project \dataset{}~\cite{OCD1M20} dataset addresses this task and how to apply it to training ML models.

If efficient approximations of DFT calculations can be performed, it opens the door to large scale exploration of new catalysts. Assuming catalysts are created from up to three of forty potential metals, over 10k combinations exist. Since each set of metals can be combined in various ratios and configurations, the total number of possibilities increases \textit{ad infinitum}. As discussed in Section \ref{sec:electrocatalyst}, determining whether a potential catalyst may be effective requires simulating its interaction with the set of potential adsorbates at all potential surfaces and binding sites. This results in O(1,000) potential simulations per catalyst. The result is a search space containing millions of possibilities requiring billions of simulations to explore! While this may seem overwhelming, with efficient DFT approximations it may be possible to compute all of them through brute force. The most promising candidates may then be verified through traditional DFT calculations before further exploration into their real-world feasibility.

While the discussion so far has focused on HES, efficient DFT approximations could be applied to numerous high-impact problems. Two notable examples include methanation and the production of ammonia fertilizer. Methanation refers to the reaction of carbon dioxide and hydrogen (generated as we described above using renewable energy) to form methane and water. Instead of storing renewable energy in the form of hydrogen this would allow storing it as methane, while having the beneficial side effect of reducing carbon dioxide (Figure \ref{fig:powercycle}). The potential end result could be a high density, carbon neutral fuel that fits into the current energy infrastructure \cite{sterner2009bioenergy}. More complex hydrocarbon fuels beyond methane, such as ethanol, may also be created through the family of Fischer-Tropsch chemical reactions. The production of ammonia, a key ingredient in agricultural fertilizer, combines hydrogen and nitrogen to create ammonia. Currently this energy intensive process, called the Haber-Bosch process, accounts for approximately $1\%$ of the world's \ce{CO2} emissions \cite{epa2016inventory}. This could be greatly reduced by using renewable energy to produce the input hydrogen and by finding a more efficient reaction to Haber-Bosch. 

\subsection{Overview}

We hope this paper serves as a gentle introduction for the machine learning community to opportunities presented by the problem of renewable energy storage, and the use of efficient ML models for atomic simulations. For researchers only interested in using the \dataset{} dataset for model training, Sections \ref{sec:storage} and \ref{sec:electrocatalyst} may be skipped, and you may proceed directly to Section \ref{sec:ml}.

We begin by providing a brief overview of renewable energy storage technologies (Sec. \ref{sec:storage}) followed by an introduction to the design of catalysts using DFT (Sec. \ref{sec:electrocatalyst}). We explore applications for deep learning starting in Section \ref{sec:ml}. This includes the \dataset{} dataset needed for training the relevant tasks (Sec. \ref{sec:datasets}) and potential directions for ML models to addressing these tasks (Sec. \ref{sec:models}). Finally, future directions are explored in Section \ref{sec:discussion}.

%% file: sections/storage.tex
\begin{table}[t]
\centering
\begin{tabular}[c]{|m{2cm}|>{\centering\arraybackslash}m{1.5cm}|>{\centering\arraybackslash}m{1.5cm}|>{\centering\arraybackslash}m{1.2cm}|}
\multicolumn{4}{c}{US Power Grid 2018} \\
\hline
Source & Capacity (GW) & Generation (TWh) & $\%$ \\ \hline
Natural Gas & 537 & 1,469 & 35.2$\%$ \\ \hline
Coal & 264 & 1,146 & 27.5$\%$ \\ \hline
Nuclear & 104 & 807 & 19.3$\%$  \\ \hline
Hydroelectric & 80 & 293 & 7.0$\%$ \\ \hline
Wind & 95 & 273 & 6.5$\%$ \\ \hline
Solar & 32 & 64 & 1.5$\%$ \\ \hline 
Other & 84 & 122 & 3.0$\%$ \\ \hline \hline
Total & 1,196 & 4,174 & -- \\ \hline
\end{tabular}
\caption{Power capacity and generation of the US power grid by source in 2018 \cite{EPA-US}.}
\label{tab:uspowergrid}
\end{table}

\section{Renewable Energy Storage}
\label{sec:storage}
In this section, we provide a brief overview of the two most popular renewable energy storage technologies---pumped-storage hydroelectricity and batteries---as well as the technologies relevant to this paper:  hydrogen energy storage and methane synthesis. A summary of these technologies is provided in Table \ref{tab:storage}. For a more thorough discussion of the numerous technologies available please see \cite{amrouche2016overview, auer2012state, 2018iv, Renewable_Storage}. Due to the readily available information provided by the US Energy Information Administration, we use the US as an example for our discussion below. However, results will vary by country based on geography, natural resources available, etc.

During this discussion we use the terms gigawatts (GW) and terawatt hours (TWh) (1 TW = 1,000 GW). A GW is a unit of measurement for electric capacity, i.e., how much electricity can flow at an instance in time. A TWh is a unit of electric energy, i.e., how much electricity was generated in one hour. A GW is equal to one billion watts and a TWh is equal to one trillion watt hours. For instance, a 50 GW plant operating for 1,000 hours at peak capacity would generate 50 TWh of energy. For reference, the largest nuclear power plant (Palo Verde) in the US has a capacity of 4 GW, while the average natural gas plant has 0.5 GW of capacity. In 2018, the US had a maximum capacity of 1,196 GW and generated 4,174 TWh of energy \cite{EPA-US}. The percentage of actual energy produced compared to the maximum capacity (known as the capacity factor) is on average $40\%$. A summary of US power generation by source is provided in Table \ref{tab:uspowergrid}.

\subsection{Estimating Needed Storage Capacity}

Estimating the total stored energy and capacity needed to move to $100\%$ renewable sources is difficult and depends on numerous factors \cite{ziegler2019storage, pump-up, nation_battery}. For illustration, we consider a case where the US experiences a week of cloudy weather in the southwest and low winds in the plains. Let's assume that during this period $70\%$ of the US energy would need to come from storage. In this case, 56 TWh of energy storage would be needed (4,174 TWh / year x 0.02 years x $70\%$) with a capacity of 420 GW (1.25 peak to mean ratio x 56 TWh / 168 hours per week). However, this assumes we're only replacing energy that is currently being supplied through electricity. When including other energy sources such as gasoline powered engines and heating, the US actually uses $\approx 30,000$ TWh in total energy \cite{MER-US}. Since electric motors are more efficient than internal combustion engines, we could potentially drop this amount by a third to 20,000 TWh. Therefore, to meet $70\%$ of our total energy needs over 7 days with storage, we would need roughly 250 TWh of storage and 2,000 GW of capacity.  Note, these numbers should only be used as very rough guides, but provide an approximate ``order of magnitude'' estimate (Table~\ref{tab:storage}) for our discussion below. 

\begin{table*}[t]
\centering
\begin{tabular}{|>{\centering\arraybackslash}m{5cm}|>{\centering\arraybackslash}m{1.5cm}|>{\centering\arraybackslash}m{1.5cm}|>{\centering\arraybackslash}m{2cm}|>{\centering\arraybackslash}m{1.5cm}|>{\centering\arraybackslash}m{1.5cm}|}
\multicolumn{6}{c}{Energy Storage Techniques} \\
\hline
Source & Capacity (GW) & Generation (TWh) & AC to AC Efficiency & Scale to Capacity? & Scale to Storage?   \\ \hline
Pumped-Storage Hydroelectricity & 23 & 24 & $70\%-80\%$ & No & No \\ \hline
Batteries & $<$1 & $<$1 & $60\%-95\%$ & Maybe & No \\ \hline
Hydrogen Energy Storage & 0 & 0 & $35\%$ & Maybe & Yes  \\ \hline
Methane Synthesis & 0 & 0 & $25\%$ & Maybe & Yes \\ \hline

\end{tabular}
\caption{Comparison of different energy storage techniques, including the installed US capacity and generation, and their round trip AC to AC efficiency. See text for detailed description. HES and methane synthesis are the only two techniques that could scale to meet both capacity and storage needs.}
\label{tab:storage}
\end{table*}

\subsection{Pumped-Storage Hydroelectricity}

\gls{PSH} refers to the conceptually simple technique of using renewable energy to pump water uphill. When additional energy is needed, the water flows back downhill through a hydroelectric plant to generate electricity. This mature technology has resulted in over 40 plants being built in the US, with most of their construction happening in the 1970s. No new plants have come online since 2012.

An advantage of PSH is their efficiency, which ranges between $70\%$ to $80\%$ for round trip AC to AC. Similar to hydroelectric plants, they are also highly responsive in that they can respond to power fluctuations in seconds. They are typically used for energy storage of 6 to 20 hours to help in system load balancing. For instance, a nuclear power plant can be kept at peak operating efficiency over 24 hours by diverting power to PSH in the night time and using hydroelectric power to meet peak demand during the daytime.

In 2018, the US PSH capacity was 23 GW \cite{EPA-US}. On average, the PSH plants operated between $8\%$ and $16\%$ of full capacity depending on the time of year (more in the summer when natural water flow into reservoirs is reduced). Assuming an average of $12\%$, the total energy produced by PSH is ~24 TWh, or $0.6\%$ percent of the power (4,174 TWh) generated by the US in 2018.

The primary limitation of PSH is its scalability. There are not enough places to store water to meet the national energy storage needs. Current yearly generation using hydroelectric is 293 TWh with a capacity of 80 GW \cite{EPA-US}, or only $7\%$ of the national total (Table \ref{tab:uspowergrid}). To meet the ~2000 GW of capacity needed during a week of cloudy still weather, we would need to build 25 times more hydroelectric dams. This would be quite a feat, especially when all the easy to build areas are already taken. A study initiated by the US Department of Energy concluded that only 65 GW of undeveloped hydroelectric potential from streams exists in the US, which wouldn't even double current capacity \cite{kao2014new}. In another US study, it was found that if every non-hydroelectric dam was converted to hydroelectric an additional 12 GW ($15\%$ of current capacity) of capacity could be created \cite{hadjerioua2012assessment}. We're also only considering capacity; this doesn't include the infrastructure needed to pump and store the water necessary to generate 250 TWh of energy. By one estimate, the needed reservoirs' area would be equivalent to the size of Lake Erie\cite{pump-up}! Even though PSH may not grow to the scale needed for nation-sized grids, PSH remains the best option for areas that do have sufficient water storage capacity.

\subsection{Batteries}

When considering electricity storage technologies, batteries quickly come to mind. Battery types can vary significantly in cost, cycle life (how many times they can be discharged and recharged), energy density, and efficiency. Similar to PSH, batteries have high efficiency ($80\%$ to $95\%$), which would seem to make them promising candidates for large scale energy storage. Of the numerous types that exist, the most popular include lead-acid, sodium-sulfur (NaS) and lithium-ion (Li-ion) batteries \cite{amrouche2016overview, Ogunniyi}. Lead-acid batteries are relatively inexpensive and have high efficiency, but have decreased in popularity due to their lower power density and limited recharge cycles ($<500$). NaS is a promising solution for large scale systems with a greater cycle life ($>1,500$) and low cost, but needs to operate at high temperatures (350 \degree C). Li-ion batteries offer a high cycle life (3,500) \cite{adachi2003development} and high efficiency ($95\%$). Due to the market for electric cars, the price of Li-ion batteries has decreased significantly, making them the lowest in per cycle cost. Currently, nearly all of the installed battery storage in the US uses Li-ion batteries for a combined capacity of 0.77 GW \cite{AEGR-US}.

Would Li-ion batteries scale to the size of the US grid \cite{Temple,nation_battery}? Recently Tesla installed a 100 MW (129 MWh) battery in Australia at a cost of \$66 million. Doing the simple math, the cost of a system with 2,000 GW of capacity would be \$1.23 trillion (other estimates put the cost at \$2.5 trillion \cite{Temple}). While this expense may be daunting, it could potentially be feasible to meet capacity demands. However, it becomes prohibitive if we consider the cost of storage. For 250 TWh of power (7 days worth of power in our cloudy low wind scenario), the costs rise to \$128 trillion! Using another estimate, the US DOE set a goal of \$100 per kWh for lithium-ion battery cells \cite{batteries2019}. Ignoring the other costs associated with batteries, this would still cost \$25 trillion. While the cost of Li-ion batteries have dropped significantly (in the last 10 years prices have dropped by ~10x), it is unlikely their cost will drop by another order of magnitude over the next several decades \cite{Cole19}. Even if cost was not an issue, there might not be enough lithium on the planet to build a battery of this magnitude\cite{nation_battery}. Finally, the cost of other rare metals used in Li-ion batteries has already significantly increased due to high demand \cite{turcheniuk2018ten}. 

As we saw in the discussion above, batteries scale better as a function of capacity than energy storage. This makes them ideal for scenarios where energy only needs to be stored for short periods. For instance, power from solar can be stored during the middle of the day to be used in the evening. Similar to PSH, this is useful in keeping more conventional plants running steadily at peak operating efficiency. Batteries also have high energy density---i.e., they take up relatively little volume---which has led to their recent popularity in electric vehicles. Unfortunately, batteries also have low specific energy---i.e., they require a large amount of mass per energy. This has prevented their use in planes, where excess mass reduces the energy efficiency of flight.

\subsection{Hydrogen Energy Storage}

Hydrogen Energy Storage (HES) refers to the process where renewable energy is stored in the form of hydrogen. Hydrogen is generated using electrolyzers powered by renewable energy to split water into hydrogen and oxygen. The hydrogen is stored for later use in fuel cells or gas-fired turbines to generate electricity. While hydrogen has historically been viewed as one of the promising technologies for moving beyond fossil fuels \cite{crabtree2004hydrogen, marban2007towards}, it is still in the early stages of adoption. When considered as an option for large scale energy storage, hydrogen offers significant potential.

Storing large scale reserves of hydrogen is feasible. Currently, there exists 4.3 trillion cubic feet of underground natural gas storage in the US using depleted gas reservoirs, salt caverns and aquifers \cite{UNG-US}. Hydrogen could be stored in the same manner in place of the natural gas. Assuming  hydrogen has 0.08 kWh of energy per cubic foot, 4 trillion cubic feet of stored hydrogen would contain 344 TWh of energy. Assuming a fuel cell efficiency of $60\%$, this corresponds to 206 TWh of stored energy; an amount very close to our target of 250 TWh of energy needed to run the US when solar and wind generation lags for a week! Moreover, if the infrastructure used to store natural gas was converted to hydrogen storage, the capital costs would be negligible.

Hydrogen is portable. It can be compressed and transported in pipes similar to natural gas for local electricity generation or transportation uses. For transport, hydrogen has more than twice the specific energy (kWh per kg) of gasoline, but $1/7$th the energy density (at 700 bar of pressure). In other words:  hydrogen storage is less efficient than fossil fuels in terms of energy stored per volume/pressure, but hydrogen is more efficient in terms of mass. This means that HES could be suitable for aircrafts. Hydrogen automobiles already have shown ranges over 300 miles (500 km).

The difficulty with hydrogen is not storage. It is in its generation and conversion back to electricity. One significant limiting factor is the $35\%$ efficiency for round trip AC to AC of HES, which is much lower than batteries and PSH. If HES is to be adopted, it needs to be less expensive than energy generated through fossil fuels. At $35\%$ efficiency, the price of renewable energy feeding HES needs to be 2.85x less expensive than energy generated through other means. Currently, the 2023 cost projection for natural gas is \$40--\$42 per MWh \cite{LECO}, which means renewable sources would need to cost \$14 per MWh. 2023 projections for wind and solar fall short of this goal with prices of \$43 and \$48 per MWh, respectively \cite{LECO}. However, the cost of renewable energy has been dropping at a high rate. For instance, the price of solar electricity has dropped from \$280 per MWh in 2010 to \$60 in 2017. The goal is to drop this to \$30 per MWh by 2030 \cite{sunshot}.

In considering HES systems, we need to consider their capital costs and not just the electricity feeding them. Using a fuel cell that is $60\%$ efficient, a kg of hydrogen will produce 20 kWh of power. Thus to generate a MWh of power we need 50 kg of hydrogen. The electrolyzer capital costs associated with producing a kg of hydrogen was estimated to be \$1.33 in 2014 and projected to fall to \$0.53 in 2025 \cite{HPC_US}. This corresponds to an additional \$26.5 per MWh in additional costs using the 2025 projections. A 250 kW fuel cell's capital cost is approximately \$300,000 \cite{battelle2016manufacturing}. With a 5-year life-cycle, this corresponds to a cost of \$27.3 per MWh, similar to the estimates of the electrolyzer. In summary, the total approximate costs for a HES system is \$113 per MWh (\$60 electricity, \$26.5 electrolyzer and \$27.3 fuel cell). For comparison, an estimate from 2009 put the total cost at \$240-280 per MWh \cite{steward2009lifecycle}. While this is still 2x to 3x the cost of electricity from natural gas, a combination of reductions in renewable energy costs and capital costs could lead to HES being price competitive.

Capital costs will begin to dominate as electricity prices decline. These costs are heavily dependent on the electrocatalysts used for both the electrolyzers and fuel cells. For instance, popular Polymer Electrolyte Membrane (PEM) fuel cell's high costs are due in part to their use of platinum as their electrocatalysts\cite{holton2013role, HFCPR_US, HPC_US, manu_cost}. A study of automotive fuel cell experts showed that $76\%$ believed the cost of platinum was the primary barrier to reducing costs \cite{whiston2019expert}. The percentage of capital costs for material platinum alone is estimated to be $8\%$ for commercial scale electrolyzers \cite{manu_cost} and $11-12\%$ for transportation fuel cells \cite{HFCPR_US}. When taking into consideration the entire cost of the electrocatalyst, which includes support structures that are needed to increase the power density of platinum, the percentage of capital costs related to platinum for fuel cells jumps to over $40\%$ \cite{HFCPR_US}.

Does there exist enough platinum to generate the needed hydrogen? We would need to generate approximately 35 TWh of power per day, assuming $70\%$ of the daily power would be generated through storage in our cloudy low wind scenario. This is equivalent to 1.75 million metric tons of hydrogen (assuming the use of a $60\%$ efficiency fuel cell). A 1~MW electrolyzer plant could produce 400 kg per day \cite{manu_cost}, resulting in the need for 4.4 million plants. Each plant uses 420 grams of platinum, for a total of roughly 2,000 metric tons of platinum. For comparison, the known platinum reserves in the world are approximately 70,000 metric tons \cite{Plat_US}. These amounts are pushing the boundaries of feasibility, especially when considering the world's energy needs and not just the US's. As we explore later in this paper, discovering new electrocatalysts that don't use rare expensive metals could lead to a significant reduction in costs, and to the potentially wide-spread adoption of HES systems. Already, research has shown that a $90\%$ reduction in the amount of platinum may be feasible \cite{ayers2016pathways,pi2015high}.

\textbf{Other methods for hydrogen generation:} Currently, the most economical process for generating hydrogen is called Steam Methane Reforming (SMR), which involves using high-temperature steam to produce hydrogen from methane. Unfortunately, a byproduct of SMR is the production of \ce{CO2} (9 kg of \ce{CO2} is produced for every kg of \ce{H2} \cite{Sun19, greenhouse2009technical}), partially defeating our goal of using hydrogen in the first place. Currently, $95\%$ of hydrogen generation in the US uses SMR \cite{eichman2016economic,usdrive2017hydrogen}. Due to the low price of natural gas, the price per kg of hydrogen from SMR is currently below \$1.50, making it competitive to gasoline \cite{usdrive2017hydrogen} (1 kg of \ce{H2} has roughly the same energy as 1 gallon of gas). Currently, 10 million metric tons of hydrogen is produced annually in the US, with $70\%$ of it used for petroleum refining and $20\%$ in fertilizer production (ammonia) \cite{CHP_US}.

\subsection{Methane synthesis}

Methane synthesis is similar to HES except energy is stored in the form of methane instead of hydrogen. Methane is the primary ingredient of natural gas ($90\%$). Methane offers two main advantages: nearly 4x higher energy density than hydrogen (by volume), and it can be used in existing natural gas infrastructure for power generation and transport. Higher energy density offers advantages in storage, especially for transportation uses.

Methane synthesis uses a process called methanation (also known as the Sabatier reaction\cite{sabatierreaction}), to combine \ce{H2} and \ce{CO2} to form methane \ce{CH4}. The input hydrogen is generated using the same electrolyzers as used in HES. The \ce{CO2} would ideally be captured from \ce{CO2} emissions from other sources, such as natural gas power plants. While capturing \ce{CO2} directly from the air is feasible, it adds significant cost due to the low concentration of \ce{CO2} in the air. The methanation process results in an additional $18-25\%$ energy loss \cite{zakeri2015electrical}, resulting in AC to AC efficiency of approximately $25\%$ \cite{kloess2012electric}.

While methanation offers a potentially carbon neutral solution to energy, it is unlikely to be adopted in the near term due to low natural gas prices. Estimated costs are approximately double that of HES \cite{kloess2012electric}. Longer-term, methanation may become economically feasible if natural gas prices increase, renewable energy and hydrogen prices decrease, or carbon taxes are levied \cite{jentsch2014optimal}. Meanwhile, methanation offers interesting challenges for research due to the more complex chemical reaction involved. Beyond methane, other even higher density and more desirable fuels may be generated, such as ethanol. Unlike methane which is a powerful greenhouse gas that is prone to leaks, ethanol is a liquid and easier to safely store. However, fuels such as ethanol have the same near-term adoption problems as methane while being even more complex to generate.

Returning to our earlier discussion on the current use of expensive, rare metals like platinum as catalysts, we next discuss the design of electrocatalysts.

%% file: sections/electrocatalyst.tex
\section{Design of Electrocatalysts}
\label{sec:electrocatalyst}

Inexpensive and effective electrocatalysts are essential for the wide-scale adoption of renewable energy storage technologies such as HES and methane synthesis. In this section, we dive into what an electrocatalyst is and their desired properties. This section is meant as an introduction to electrocatalysis for the non-expert that may want to experiment with building machine learning models to predict various molecular proprieties. For those interested in a more detailed exposition, please refer to \cite{Norskov14}.

\subsection{Electrocatalyst Properties}

An electrocatalyst is a catalyst used to increase the reaction rate and efficiency of an \gls{electrochemical_reaction} (a chemical reaction that generates electricity). As is true for all catalysts, an electrocatalyst is not consumed during the reaction. In the design of an catalyst, we must consider several properties to determine whether it will be useful in a practical scenario. These include:
\begin{itemize}
    \item {\bf Efficiency} - What fraction of energy is lost when driving the reaction?
    \item {\bf Reaction Rate} - How fast does the reaction occur? Typically measured in moles per second.
    \item {\bf Stability} - How durable is the catalyst, i.e., how long will it last?
    \item {\bf Selectivity} - Does it generate only the desired products, or other byproducts?
    \item {\bf Cost} - How expensive is it to manufacture the catalyst?
\end{itemize}
In this paper, we will primarily discuss the efficiency and reaction rate of an catalyst. The intent of focusing on these characteristics is to screen cheaper catalysts by their efficiency and reaction rates. However, the other factors are still very important. The stability of an catalyst determines whether it will degrade during prolonged use in a reaction. If an catalyst breaks down, its costs will significantly increase due to the need to replace it. Selectivity refers to the percentage of the desired product produced by the reaction. In many reactions, unwanted byproducts may be produced. While we don't discuss selectivity specifically, similar tools to those used to determine a reaction's efficiency and reaction rates can also be used to calculate its selectivity. Finally, the manufacturing cost of the catalyst must be taken into account. This is driven by two primary factors:  the cost of the materials used in its construction, and the complexity of the structure. Ideally, an catalyst should be constructed from abundantly available inexpensive elements, and not platinum, iridium, etc.

\begin{figure}
	\begin{center}
		\includegraphics[width=0.92\linewidth]{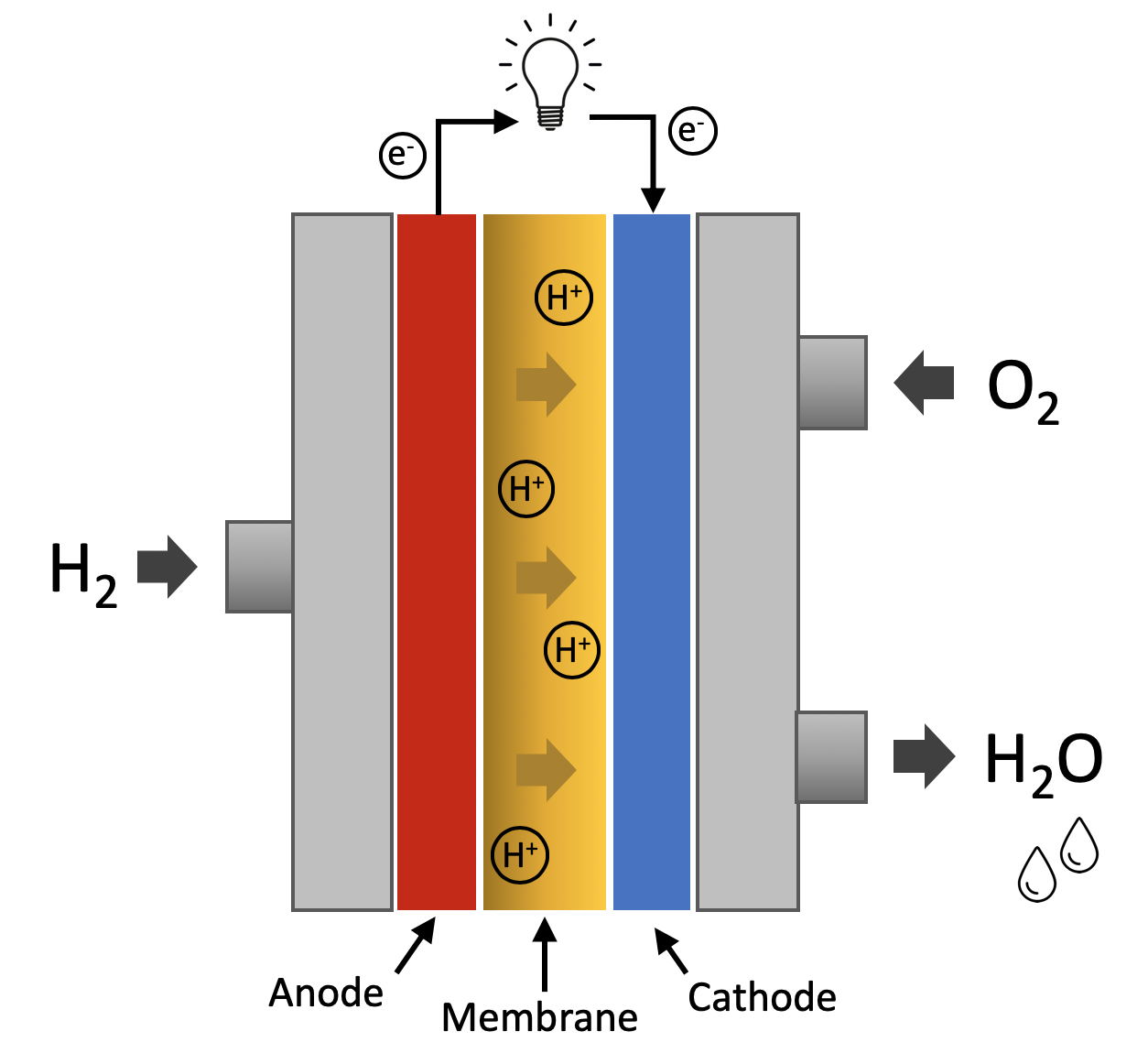}
	\end{center}
	\caption{Illustration of a PEM fuel cell. The anode (red) splits hydrogen into protons \ce{H+} and electrons \ce{e-}. The cathode (blue) combines the protons and electrons with oxygen to produce water. The membrane (yellow) only allows protons to pass through it, which forces the electrons to travel through the electric circuit, creating an electric potential.}
	\label{fig:FC}
\end{figure}

\subsection{Fuel Cells}
To ground our discussion, we begin by describing fuel cells, which are a specific use case for catalysts. While we use this as an illustrative example, other electrochemical reactions behave in a similar manner, such as electrolyzers.

A fuel cell is a conceptually simple device that converts hydrogen and oxygen into water and electricity. A fuel cell contains three main parts; the anode, cathode, and membrane (Figure \ref{fig:FC}). In the popular Proton-Exchange Membrane (PEM) fuel cell, the anode breaks the hydrogen into protons \ce{H+} and electrons \ce{e-}:
$$\ce{H2 -> 2H+ + 2e-}$$
The cathode combines oxygen \ce{O2} with the protons and electrons from the anode reaction to create water \ce{H2O}:
$$\ce{1/2O2 + 2H+ + 2e- -> H2O}$$
Taken together, the overall chemical reaction is:
$$\ce{H2 + 1/2O2 -> H2O}$$
How does a chemical reaction that combines hydrogen and oxygen to create water also generate electricity? The secret is in the membrane. The membrane only allows protons to pass through it, which forces the electrons to take an alternative path through the electrical circuit generating electricity!

The amount of electricity generated is dependent on how fast the reactions on the anode and cathode occur. As we discuss next, the rate of these reactions is highly dependent on the type of catalysts used.

\subsection{Electrocatalysis}

A catalyst is a substance used to increase the rate of a chemical reaction that is not consumed in the process. A chemical reaction pathway typically has several intermediate steps. The catalyst works by providing an alternative reaction pathway that reduces the amount of energy, voltage, or pressure needed for the reaction to occur. There are three main types of catalysts:  homogeneous catalysts that are in the same phase as the reactant (typically gas-gas or liquid-liquid), heterogeneous catalyst that are in a different phase (typically gas-solid), and enzymes that are biological catalysts. We focus on heterogeneous catalysts, since they allow for the easy separation of the catalyst (solid) from the reactants (gas) in practical applications.

An electrocatalyst is a catalyst used during a electrochemical reaction. In our fuel cell example, both the anode and cathode would use a catalyst to increase their reaction rates. To understand how this is done, let us consider the Oxygen Reduction Reaction (ORR) that occurs at the fuel cell's cathode, which involves the splitting of \ce{O2} to create water \cite{norskov2004origin, Khotseng18,kulkarni2018understanding}:
\begin{equation}\ce{1/2O2 + 2H+ + 2e- -> H2O}\label{eqn:cathode}\end{equation}
Most reactions proceed along a pathway containing a series of smaller steps. For any given reaction there may even be multiple possible pathways, e.g., Figure \ref{fig:pathway} for syngas. One potential reaction pathway (the dissociative pathway \cite{norskov2004origin}) for (\ref{eqn:cathode}) involves three intermediate steps\footnote{The higher complexity associative pathway containing five steps is more likely to be important for platinum catalysts \cite{kulkarni2018understanding}}:
\begin{equation}\ce{1/2O2 + {*} -> {\mbox{*}}O}\label{eqn:step1}\end{equation}
\begin{equation}\ce{{\mbox{*}}O + H+ + e- -> {\mbox{*}}OH}\label{eqn:step2}\end{equation}
\begin{equation}\ce{{\mbox{*}}OH + H+ + e-  -> H2O + {*}}\label{eqn:step3}\end{equation}
where * indicates a bonding site on the catalyst. In (\ref{eqn:step1}), oxygen gas \ce{O2} comes in contact with the catalyst's surface which splits apart the two oxygen atoms. While still on the surface of the catalyst, an oxygen atom O{\mbox{*}} combines with a proton \ce{H+} and electron \ce{e-} to form \ce{OH} in (\ref{eqn:step2}). Finally in (\ref{eqn:step3}), another proton and electron combine with the \ce{OH} to form a water molecule. The water molecule has a weaker bond with the catalyst, allowing it to float away. Why is the catalyst needed? Without the presence of a catalyst, breaking the bond between the pair of oxygen atoms and keeping them separated so that they can bond with the protons and electrons would require a significant amount of energy (heat). When it comes in contact with the catalyst, the bond between the oxygen atoms is weakened as they bond with the catalyst. This allows the rate of the reaction to increase with less energy.

\begin{figure}
	\begin{center}
		\includegraphics[width=0.92\linewidth]{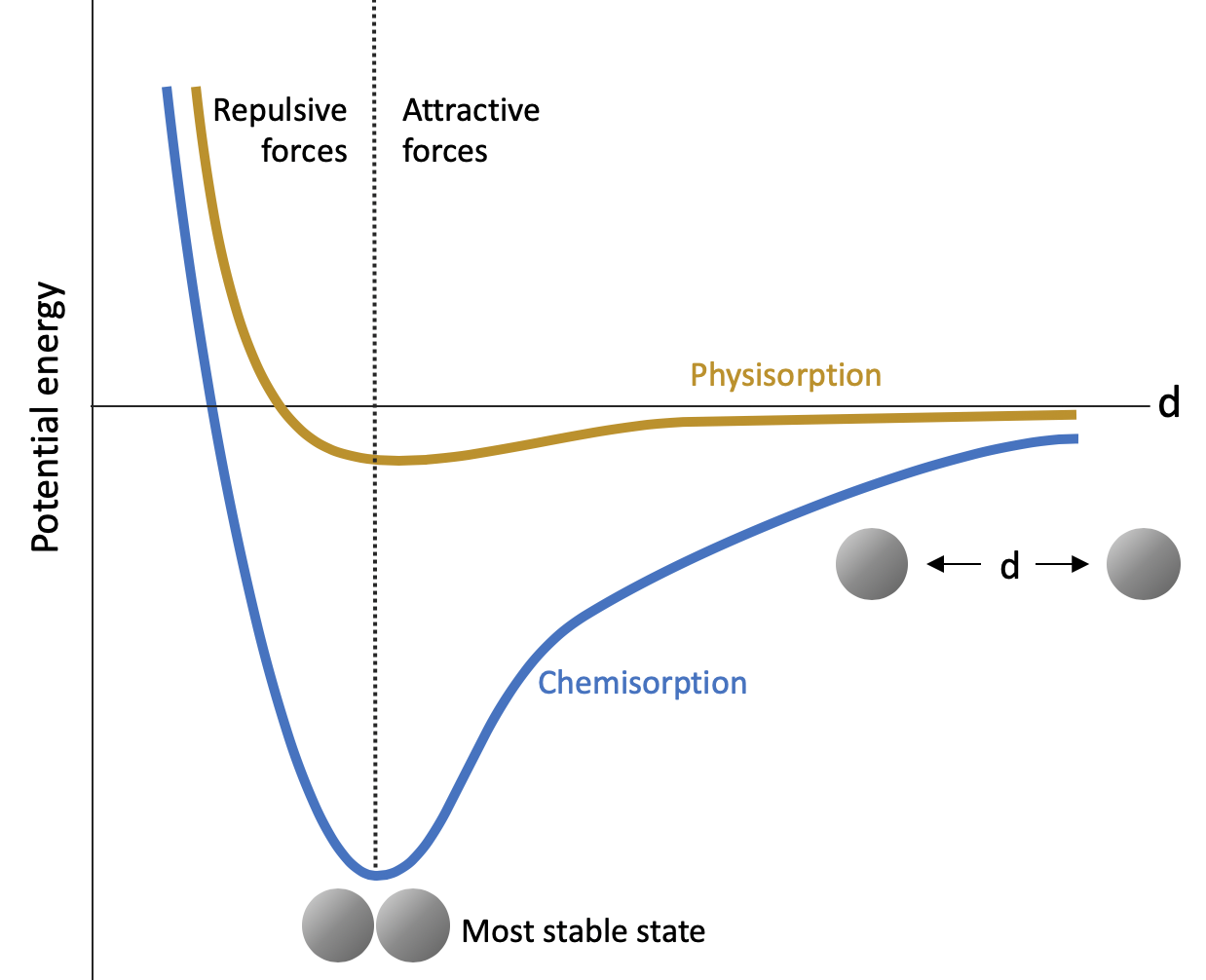}
	\end{center}
	\caption{Illustration of the potential energy between two atoms as the distance $d$ between them varies. When the atoms are far apart (right), the weaker van der Waals forces attract atoms towards each other. As the atoms move closer together (left), the stronger Coulombic forces repel the atoms. The minimum is the stable state where these two forces cancel out. The depth of the minimum is dependent on whether a bond is formed (chemisorption) or not (physisorption) by the atoms.  }
	\label{fig:PED}
\end{figure}

To understand this process in more detail, we need to define a few terms. The amount of usable energy for a closed system with constant temperature and pressure is referred to as the \gls{GFE}, or Gibbs energy and is defined as:
\begin{equation}
    G = H - TS
    \label{eqn:gfe}
\end{equation}
where $H$ is the \gls{enthalpy} (energy contained in the bonds between atoms) of the system, $T$ is the temperature (measured in Kelvin) and $S$ is the entropy. The entropy of a system increases when molecules break their bonds and decreases when they form new ones. The computation of $H$ involves the potential energy between atoms, which is illustrated for two atoms as a function of distance in Figure \ref{fig:PED}. When the two atoms are far apart, they are weakly attracted to each other. As they move closer together this attraction is maximized resulting in a minimum potential energy. If the atoms move even closer together, they begin to repel and the potential energy increases. If a bond is formed between two atoms (chemisorption), their attraction is stronger and the minimum potential energy is lower than if a bond is not formed (physisorption). The kinetic energy measures the energy due to the motion of the system, i.e., how fast are the atoms moving or simply the heat of the system.

An important property in determining whether a reaction is spontaneous (whether it will occur without energy input) is its change in Gibbs energy from the initial state to the final state. The change in Gibbs energy is referred to as the ``free energy'' for a reaction, and is defined as:
\begin{equation}
    \Delta G = \Delta H - T\Delta S
\end{equation}
If the difference in Gibbs energy is negative the system is said to be \gls{exergonic}, i.e., it will occur spontaneously (similar to a cart being pulled down a hill by gravity). Otherwise the reaction is \gls{endergonic} with the Gibbs energy increasing. By its definition, an endergonic reaction requires energy to be added to the system for it to occur (the cart needs to be pushed up the hill). For scenarios in which the temperature is relatively low, the \gls{free_energy} will be dominated by the change in enthalpy ($\Delta H$ term). A negative $\Delta H$ means heat is released by the reaction (\gls{exothermic} reaction), where as a reaction that absorbs heat has a positive $\Delta H$ (\gls{endothermic} reaction).

\begin{figure}
	\begin{center}
		\includegraphics[width=0.92\linewidth]{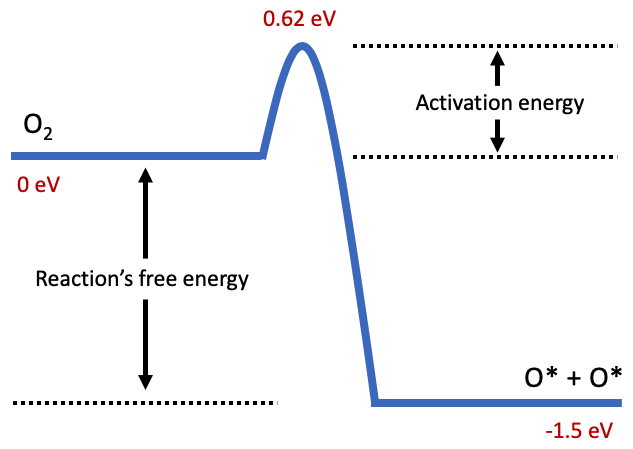}
	\end{center}
	\caption{Illustration of a reaction's free  and activation energy. The primary role of a cataylst is to minimize the activation energy of a reaction. The sign of the free energy indicates whether the reaction is spontaneous (negative) or not (positive). Plot values obtained from CatApp \cite{hummelshoj2012catapp} for Pt(111) surface.}
	\label{fig:O2}
\end{figure}

At first glance, it may be tempting to assume that all exergonic or exothermic reactions will spontaneously occur at a high rate. After all, any system will want to reduce its potential energy much like a cart let go on a hill. However, this isn't always true. In many cases, a set of atoms will be bonded together, and be quite satisfied in their current state, such as the \ce{O2} gas molecule in reaction (\ref{eqn:step1}). In terms of our analogy:  The cart could have a wedge under one of its wheel to prevent it from rolling, even though it may ``want to'' roll downhill. In order for the oxygen atoms to bond with the catalyst, they first need to break their oxygen-oxygen bonds. Breaking this bond requires kinetic energy, and the amount of energy needed is referred to as the \gls{activation_energy} for the reaction, Figure \ref{fig:O2}. In our cart analogy:  This is how much effort it takes to remove the wedge from under the cart's wheel. As an extreme example, the decay of diamonds to graphite is a exergonic reaction, but due to its very high activation energy this process takes millions of years to occur!

We may think of each state in the reaction as a local minimum in the Gibbs energy with the activation energy being the height of the hills between the valleys. The primary goal of a catalyst is to reduce the activation energy, or the height of the hills between the valleys. As shown in Figure \ref{fig:O2}, the free energy between \ce{O2} and \ce{2O\mbox{*}} in (\ref{eqn:step1}) is -1.5eV with a platinum catalyst (eV is a unit of energy measurement that is commonly used for chemical reactions). Even though the reaction is clearly exothermic, the activation energy of reaction (\ref{eqn:step1}) is still 0.62 eV to break the \ce{O2} bond. Therefore, energy (heat) needs to be added to the system in order for the reaction to occur, even though the final potential energy is lower than the initial energy. Note that in a real system, the activation energy does not act as an on/off switch. Instead, the rate of reaction gradually increases as the activation energy is approached.

\begin{figure}
	\begin{center}
		\includegraphics[width=0.92\linewidth]{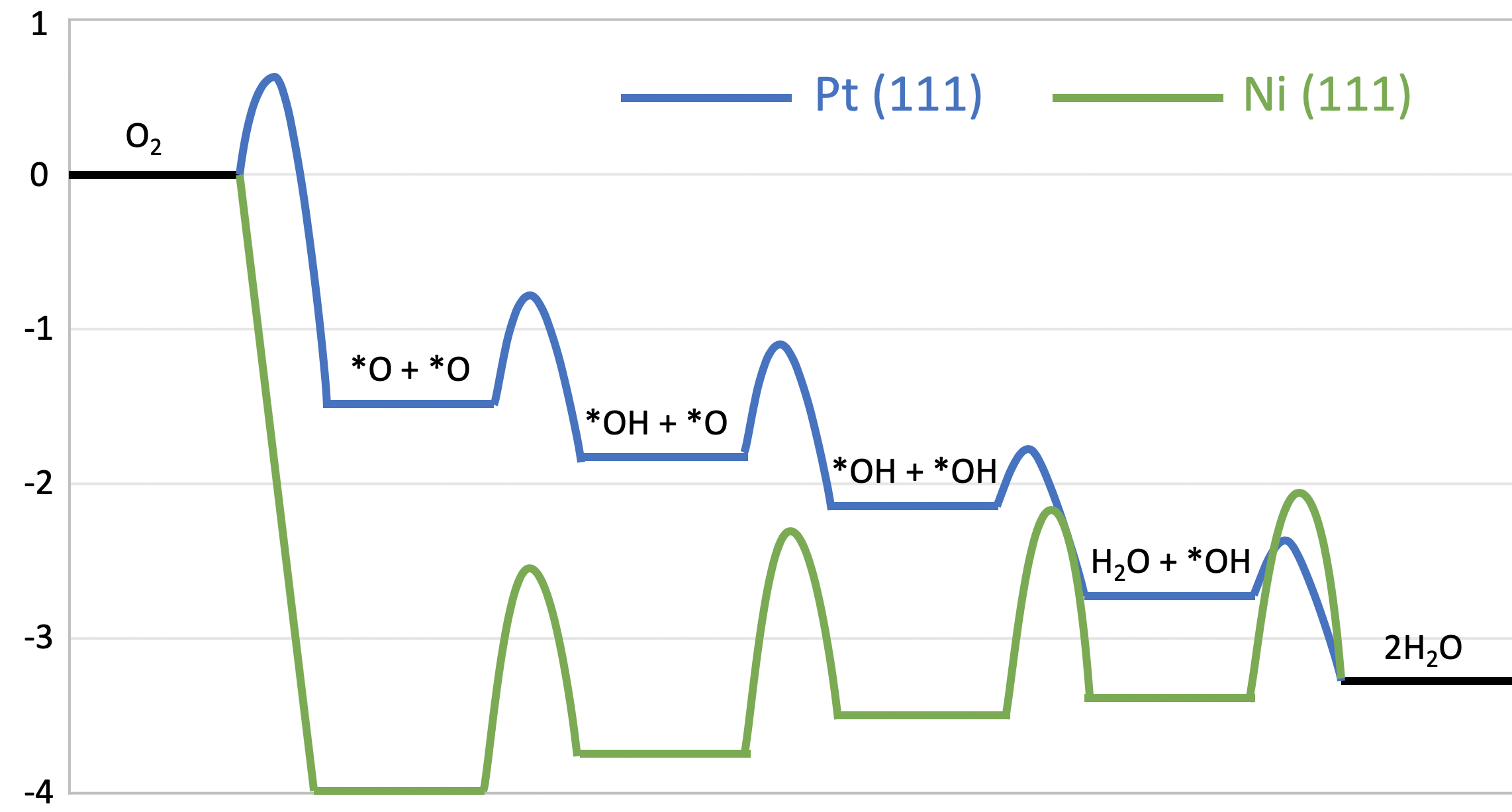}
	\end{center}
	\caption{Illustration of Gibbs free energy diagram for platinum (111) and nickel (111) catalysts for the dissociative pathway of reaction (\ref{eqn:cathode}). Activation energies are approximated from equivalent thermal reactions.} 
	\label{fig:cathode}
\end{figure}

The cathode's oxygen reduction reaction (\ref{eqn:cathode}) contains three intermediate steps. To get a complete picture of the overall reaction, we can string together Gibbs free energy plots for the three \gls{intermediate} reactions to obtain Figure \ref{fig:cathode} (two steps are repeated twice to create two water molecules). The figure shows the reaction in the presence of two different catalysts, platinum and nickel. The overall reaction is exothermic using either catalyst. However, platinum has a much faster rate of reaction. Why is this? Note that the reaction's overall free energy is independent of the catalyst used. The choice of catalyst only impacts the free energy and activation energies of the intermediate reactions. Using platinum, most of the intermediate reactions have lower activation energies. Using nickel, the oxygen atoms bind very strongly with the catalyst, resulting in a significantly lower free energy. In fact, no activation energy is needed to split \ce{O2} in the first reaction. However due to the strong oxygen binding, the second two reactions have high activation energies, which slows the overall reaction rate. 

\begin{figure}
	\begin{center}
		\includegraphics[width=0.92\linewidth]{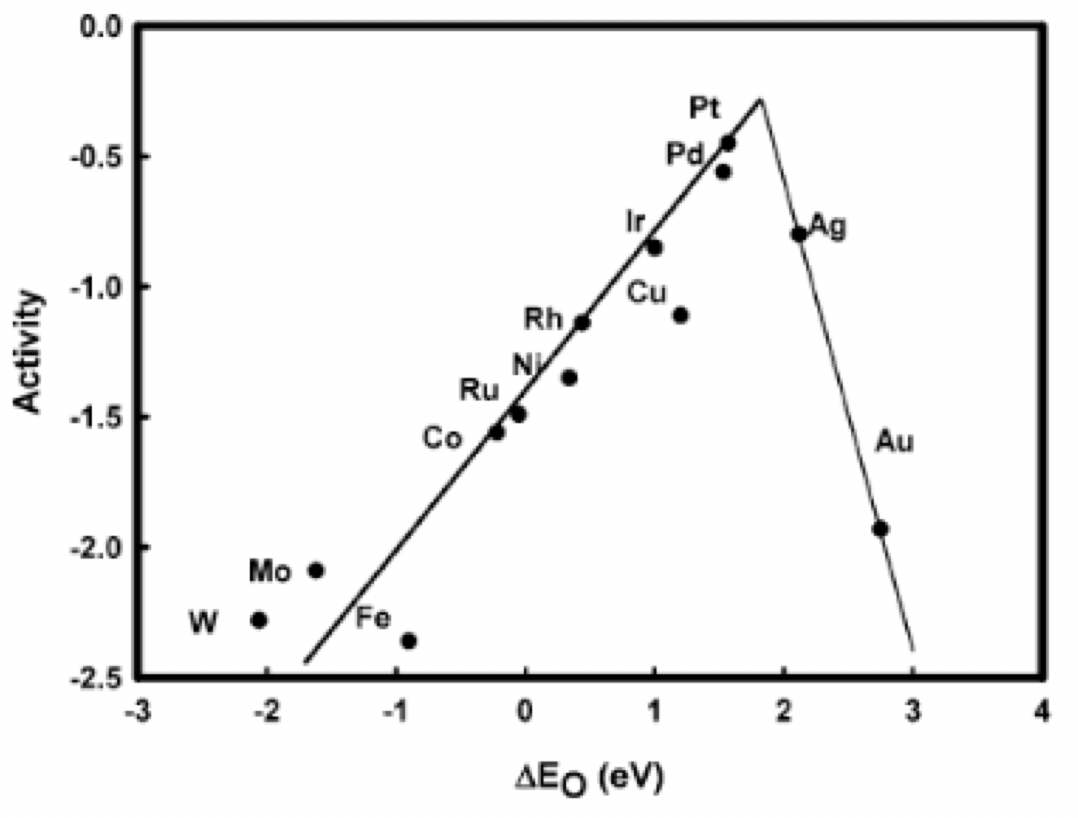}
	\end{center}
	\caption{Plot of reaction activities or rates (vertical axis) vs. the oxygen adsorption energy $\Delta E_O$ for different metals. Lines indicate the ``volcano plot'' using Sabatier analysis to predict the reaction rates. Note that expensive metals such as platinum (Pt) and palladium (Pd) are near the ideal peak, where as the more common metals iron (Fe) and nickel (Ni) have significantly lower rates. Reprinted with permission from \cite{norskov2004origin}. Copyright 2004 American Chemical Society.}
	\label{fig:metals}
\end{figure}

Figure \ref{fig:metals} shows the reaction rates of various metals as a function of the binding energy between oxygen and the catalyst. Note the reaction rate is like Goldilocks. If the binding energy is too low (strong bond) the oxygen has a hard time reacting and separating from the catalyst (nickel Ni, iron Fe). On the other hand, if the binding energy is too high (weak bond), the catalyst (silver Ag or gold Au) has a hard time attracting and holding onto oxygen atoms so that the desired reactions can take place. Platinum conveniently sits near the peak, with the ``just right'' bonding energy.

\begin{figure}
	\begin{center}
		\includegraphics[width=0.92\linewidth]{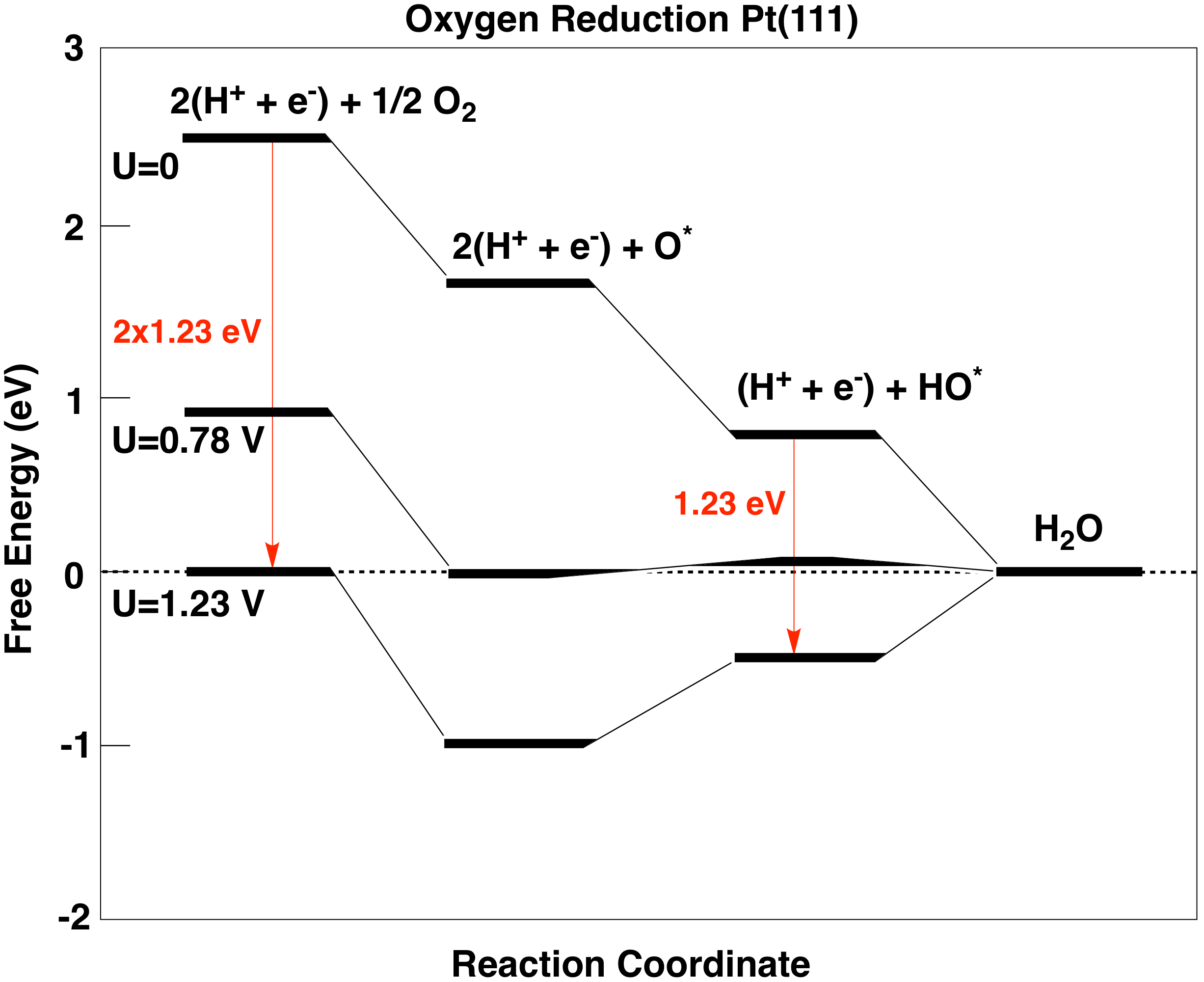}
	\end{center}
	\caption{Illustration of Gibbs free energies for the dissociative pathway of the ORR reaction (\ref{eqn:cathode}) with adjustments based on different electric potentials $U$. These include zero potential ($U=0V$), the highest potential where all steps are exothermic ($U=0.78V$), and the equilibrium potential ($U=1.23V$). Adapted with permission from \cite{norskov2004origin}. Copyright 2004 American Chemical Society.}
	\label{fig:elec}
\end{figure}

In the discussion above, we've considered the fuel cell system as if it was short-circuited, i.e., no resistance was placed on the electronic current. In practice, the circuit will have resistance, e.g., to turn on a light bulb. What effect does this have on the Gibbs free energy? Resistance in the circuit between the electrodes creates potential energy that can also be measured in volts. To find the resulting free energy given different electrode potentials, we can just subtract the potential energy per electron for each step in our reactions. As shown in Figure \ref{fig:elec}, the equilibrium potential (when the overall free energy is zero) is 1.23 eV, and the highest potential in which all reactions are exothermic is 0.78 eV. Thus the free energy of an electrochemical reaction not only helps determine its spontaneity, but also the electric potential that may be generated.

\subsubsection{Reaction rate}

The rate of a chemical reaction typically follows the form:
\begin{equation}
    r = k[A][B]
\end{equation}
where $r$ is the rate of the reaction, $k$ is the ``rate constant'', $[A]$ is the concentration of one reactant, and $[B]$ is the concentration of the other reactant. Thus the reaction rate is dependent on two types of quantities:  the concentrations of the reactants (the inputs to the reaction) and the reaction's rate constant. Intuitively if the concentration of the reactants is higher, the reaction rate will increase. The reaction rate constant $k$, for which the reaction rate varies linearly, is computed using the Arrhenius equation that contains several terms:
\begin{equation}
    k = Ae^{-E_a/RT}
\end{equation}
where the constant $A$ encodes how frequently the reactant molecules collide and whether they have the correct orientation to react. The second term $e^{-E_a/RT}$ measures the probability of a collision having enough energy (velocity of the molecules) for a reaction to occur. 

It is in this second term that the activation energy $E_a$ is needed. $E_a$ is divided by the temperature $T$ and the gas constant $R$. Thus, the higher the activation energy, the higher the temperature needed to drive the reaction. If an catalyst lowers the activation energy, the reaction rate will increase for the same temperature, increasing the reaction rate.

\subsubsection{Estimating reaction rates in practice}
\label{sec:rates_in_practice}
There are many factors that impact the overall rate, such as temperature, pressure, coverage (percentage of binding sites occupied), surface area of the catalyst, and the rate constant. However, there are some methods for changing the rate constant that extend beyond the relatively simple Arrhenius equation, such as changing the electrolytes used in the electrochemical cell or using promoters \cite{wang2018electrolyte,vayenas2001electrochemical} (additional non-catalytic substances used to increase the reaction rate). Given all these factors it can be very difficult to directly compute the free and activation energies needed to compute the rate constant and reaction rate. However, it has been shown experimentally that the variation in the reaction rate can be approximated using only a small number of factors.

In most reactions there is a single reaction step that limits the rate of the reaction---i.e., a bottleneck reaction step. Thus only this intermediate reaction needs to be considered for determining the overall reaction rate. An intermediate reaction can be described using the change in Gibbs free energy (free energy) and the activation energy. The Gibbs free energies are computed in part (during the computation of $\Delta H$) based on the \gls{adsorption_energy} of the adsorbate on the catalyst, i.e., the change in energy of the system before and after the adsorbate bonds with the catalyst. Perhaps surprisingly, it has been shown that adsorption energies and activation energies are related by a linear function across a wide range of potential catalysts \cite{bronsted1928acid,norskov2002universality}. For instance, consider reaction (\ref{eqn:step1}) in our oxygen reduction reaction. If a catalyst bonds more strongly to the individual oxygen atoms, it is also more likely to be able to pull apart the oxygen atoms in \ce{O2}. The linear relation between energies results in the the ability to reasonably approximate the overall reaction rate using a linear function to predict activation energies from the adsorption energies of the intermediate adsorbates.

\begin{figure}
	\begin{center}
		\includegraphics[width=0.92\linewidth]{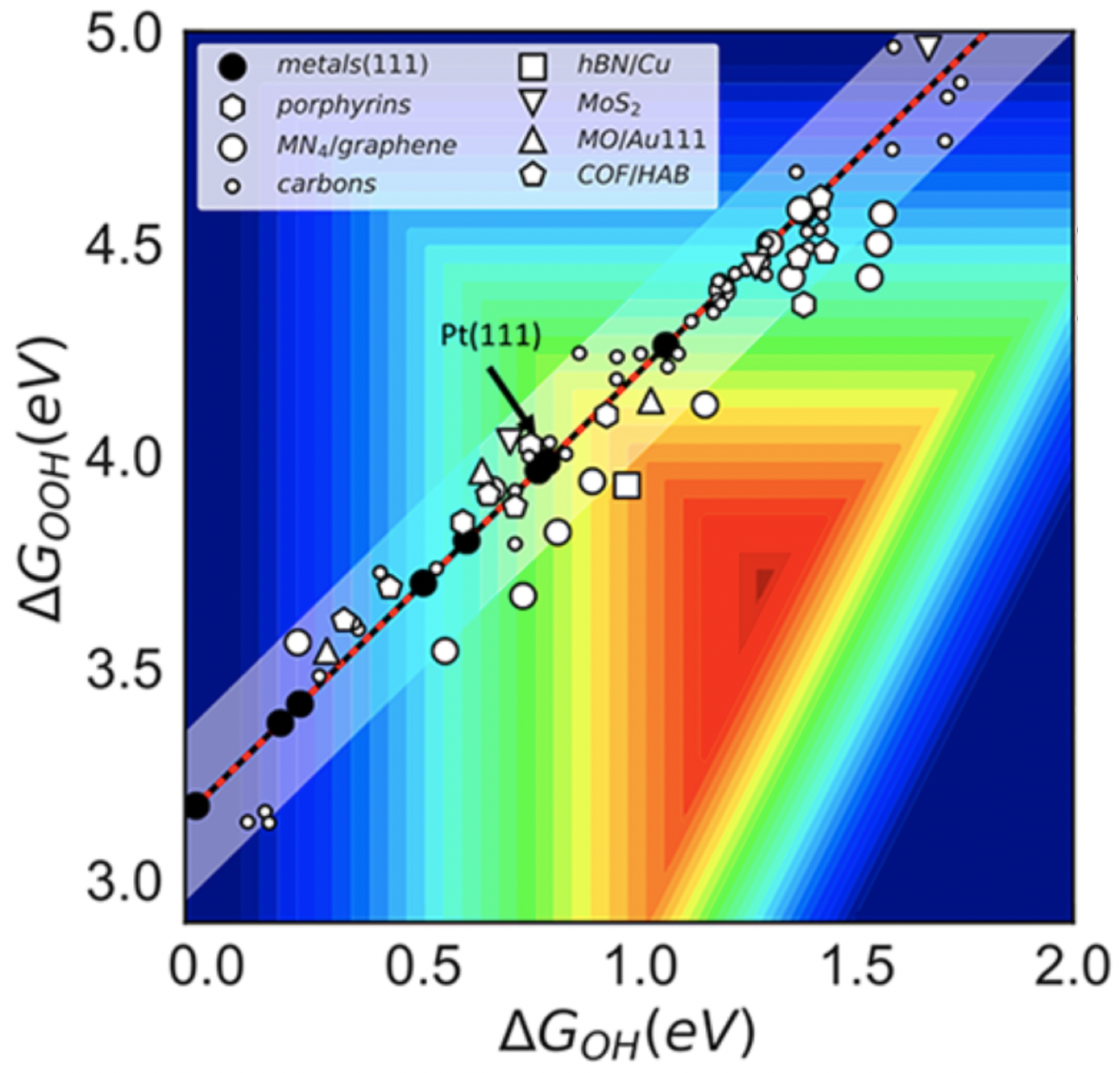}
	\end{center}
	\caption{Plot showing reaction activity (red higher, blue lower) for the ORR reaction (\ref{eqn:cathode}) using the associative pathway. Plotted as a function of the binding free energy of \ce{{\mbox{*}}OH} and \ce{{\mbox{*}}OOH}. Note that most known catalysts lie along a line due to the high correlation in binding or adsorption energies between \ce{OH} and \ce{OOH}. Finding a catalyst closer to the red peak remains an open research problem. Reprinted with permission from \cite{kulkarni2018understanding}. Copyright 2018 American Chemical Society.}
	\label{fig:metals2D}
\end{figure}

Depending on the catalyst, the rate limiting reaction will change. Using a technique called Sabatier Analysis \cite{sabatier1986top}, the various linear relations for each rate limiting reaction can be combined by taking their minimum. The end result are plots, sometimes referred to as ``volcano plots'', like those shown in Figures \ref{fig:metals} and \ref{fig:metals2D} from \cite{norskov2004origin} and \cite{kulkarni2018understanding} respectively. In Figure \ref{fig:metals}, the two lines signify two different rate limiting reactions: combining \ce{O} and \ce{H+} (left) and separating \ce{O2} (right). The point indicates the transition between the two, and the optimal adsorption energy ($\sim 1.8$eV) in reaction (\ref{eqn:step1}). A similar plot is shown in \ref{fig:metals2D} for a different reaction pathway (associative vs. dissociative) for reaction (\ref{eqn:cathode}), but in two dimensions using the energies for two intermediate adsorbates (\ce{OH} and \ce{OOH}). Using these approximations, the reaction rates of new catalysts can be optimized by searching for the appropriate adsorption energies per adsorbate.

It is interesting to note that for a wide range of different catalysts, they all lie roughly on a line in Figure \ref{fig:metals2D}. Why is this? The reason is the adsorption energies of \ce{OOH} (vertical axis) and \ce{OH} (horizontal axis) to a catalyst are highly correlated, since \ce{OH} and \ce{OOH} bond through their \ce{O} atoms and not their \ce{H} atoms \cite{kulkarni2018understanding}. Essentially, the catalysts have a very difficult time distinguishing between these two molecules. This places limits on the types of catalyst that are likely to be found, since it is highly unlikely that a new catalyst will deviate from this trend. In order to differentiate between \ce{OH} and \ce{OOH}, research into more complex structures is needed \cite{kulkarni2018understanding}.

Up until this point, we've assumed that the Gibbs, adsorption, and activation energies, which are important for determining a reaction's spontaneity and rate, are known. How are these values computed in practice? Determining these values experimentally can be difficult, since most chemical reactions are comprised of multiple potentially unknown steps. Instead, these values are computed through numerical approximation from first principles. A popular technique for this is called Density Functional Theory (DFT) \cite{sholl2011density, parr1980density}. As we explore next, DFT can produce accurate results for problems like those discussed above, but the computational cost of DFT is high. Which raises the central question of this paper:  Can we train deep learning models to efficiently approximate DFT and enable the large-scale exploration of new catalysts?

\subsection{Density Functional Theory}

Density Functional Theory (DFT) is a popular quantum chemistry method for computing the electronic properties of an atomic structure including its energy and per-atom forces \cite{parr1980density}. Here, a structure refers to a set of atoms and their 3D positions, which are input and assumed to be fixed. In our scenario, the structure includes the atoms of the adsorbate and the catalyst. Adsorbates are molecules that participate in the chemical reaction of interest, e.g., \ce{O}, \ce{HO} and \ce{H2O} in the oxygen reduction reaction (\ref{eqn:cathode}). 

Using the energies computed at multiple structures, we can determine the reaction's free, adsorption, and activation energies. For instance, we can determine the adsorption energy by taking the energy difference between a structure where the adsorbate is in contact with the catalyst and a structure where the adsorbate is far away from the catalyst. A reaction's free energy can be computed by taking the difference in energy between the final molecule and the initial molecule along with some additional corrections. Finally, if the transition state (the state corresponding to the peak in energy between two reaction steps) of a system is known, the activation energy can be computed by finding the energy difference between the reaction's initial molecule structure and the transition structure.

DFT takes as input the positions of the structure's atomic nuclei and computes its energy using the electron density \cite{sholl2011density}. The energy of a structure, which is known as its \gls{electronic_energy}, refers to the potential energy, kinetic energy and other energies related to the interactions of the electrons \cite{baseden2014introduction}. While we assume in our experiments that the atom nuclei are stationary (no kinetic energy), the electrons move and have both kinetic and potential energy. The energy is found using an approach that iteratively estimates the electron density of the structure until convergence. The electron density is represented using a set of spatially dependent functions, which may be either spatially localized for isolated molecules, or periodic functions for extended surfaces. This process is made possible using the Kohn-Sham equations\cite{kohn1965self}. Note the term density functional theory comes from the use of functionals of the electron density, i.e., functions of the electron densities that are represented as functions. 

\subsubsection{Relaxations}

When estimating the energy of an adsorbate interacting with a catalyst, the positions of the atoms' nuclei are typically only roughly known. If we used DFT to calculate energies using these rough estimates, the predicted energies could be significantly overestimated, since real-world systems seek the state of lowest energy. Instead, the energy should be computed using the ``relaxed'' structure of the system. That is, the structure in which the atomic positions have been optimized to find the configuration with the lowest energy. The relaxation process is done iteratively using second order optimization methods. The initial atom positions are determined through heuristics based on the expected bonds. Assuming the nuclei positions are fixed, DFT is used to compute the electron densities and energy. Using derivatives of the structure's energy, the forces on the nuclei are found and their positions updated. This process is repeated until the nuclei positions converge and the minimum energy is found. Typical relaxations require 50--400 iterations to converge. By making thermodynamic corrections (such as changes, which we don't describe, from the $TS$ term in Eqn.(\ref{eqn:gfe})) to the resulting energy, the structure's Gibbs free energy can also be computed. For a more detailed explanation for how the energies and electron densities are computed, please refer to \cite{sholl2011density}.

Unfortunately, there exist limitations to the use of DFT. Most importantly, it is computationally expensive and scales $O(n^3)$ with respect to the number of electrons in the structure. In practice, this results in DFT relaxations taking hours--weeks per simulation using O(10--100) core CPUs on structures containing O(10--100) atoms. As a result, complete exploration of catalysts using DFT is infeasible. DFT relaxations also fail more often when the structures become large (number of atoms) and complex. It is not uncommon for relaxations to not converge, especially if the initial structure is not set properly. In practice, failure rates vary from $20\%$ to $50\%$ for the types of structures we consider in this paper. DFT would fail completely for large and more complex catalysts such as enzymes. The hope is that machine learned approximations to DFT can be both faster and potentially more stable enabling the exploration of more and larger structures.

\begin{figure}
	\begin{center}
		\includegraphics[width=0.92\linewidth]{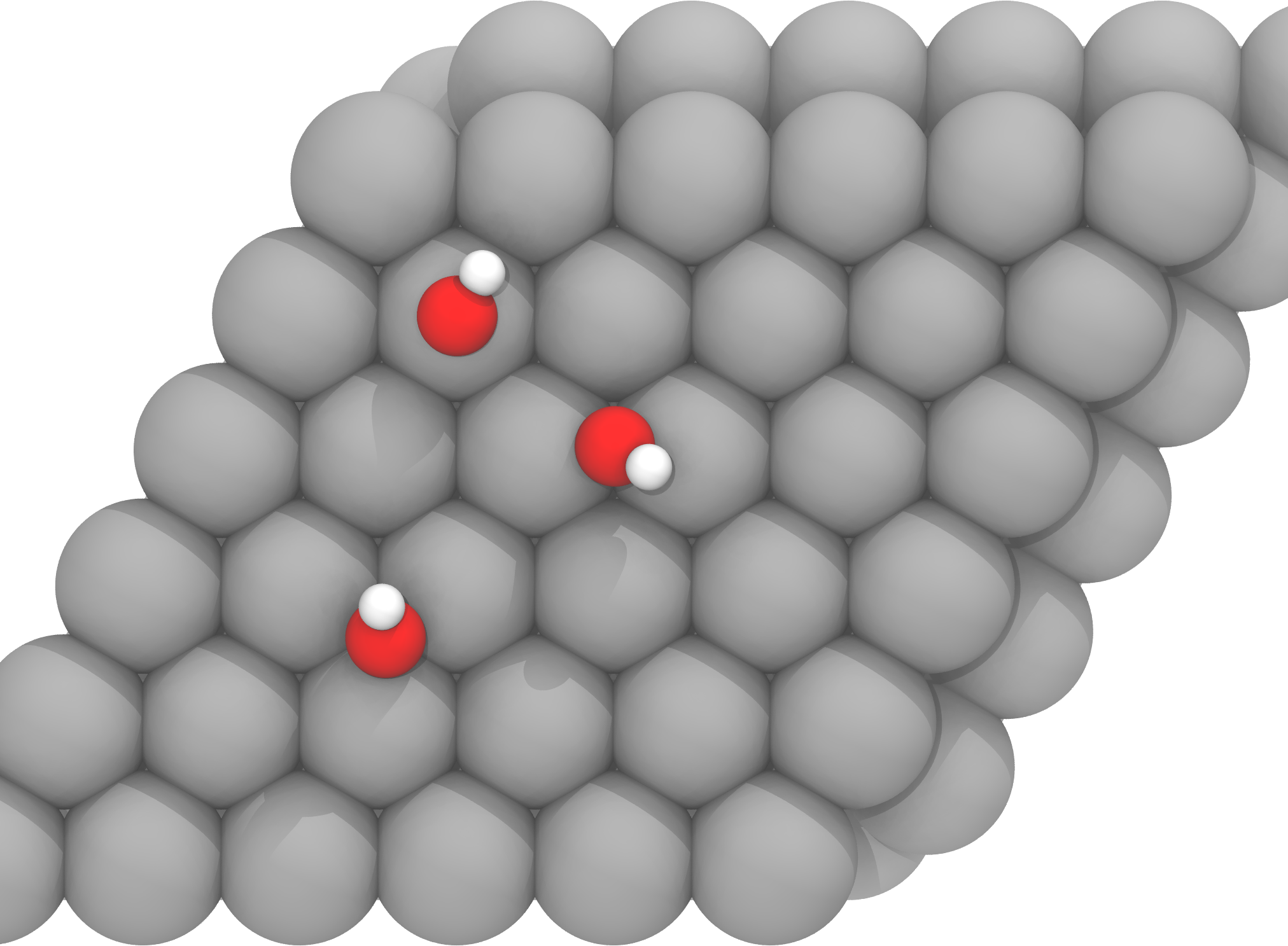}
	\end{center}
	\caption{Illustration of three binding sites for \ce{OH} on a metallic surface. The oxygen atom is shown bonding with one (top), two (middle) and three (bottom) metal atoms.}
	\label{fig:binding}
\end{figure}

\subsubsection{Initial Structure}

Before a relaxation is run, an initial structure needs to be defined. In this paper, we consider structures containing three attributes: a catalyst, an adsorbate, and a binding site. Other attributes, such as temperature and pressure, will be held constant. The catalyst is assumed to have a crystalline structure, i.e., a structure containing a repeating pattern called a ``slab''. To make the DFT computations feasible, the catalyst is modeled using a single ``slab'' that is tiled to create the catalyst's 2D surface. A typical slab is 6 atoms deep and 4-6 atoms across. The adsorbate is selected from the set of molecules in the reactions of interest and their intermediates.

\begin{figure*}
	\begin{center}
		\includegraphics[width=0.92\linewidth]{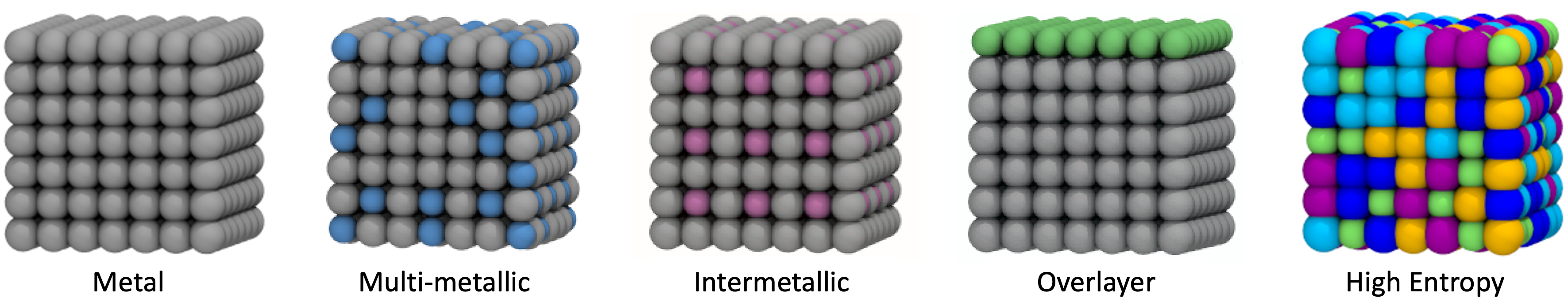}
	\end{center}
	\caption{Illustration of different metallic catalyst bulk structures: pure metals, multi-metallic (1-3 metals), intermetallic (ordered multi-metallic), overlayer (thin layers of metal applied to the surface), and high-entropy (more than 3 metals).}
	\label{fig:metalsurfaces}
\end{figure*}

\begin{figure}
	\begin{center}
		\includegraphics[width=0.92\linewidth]{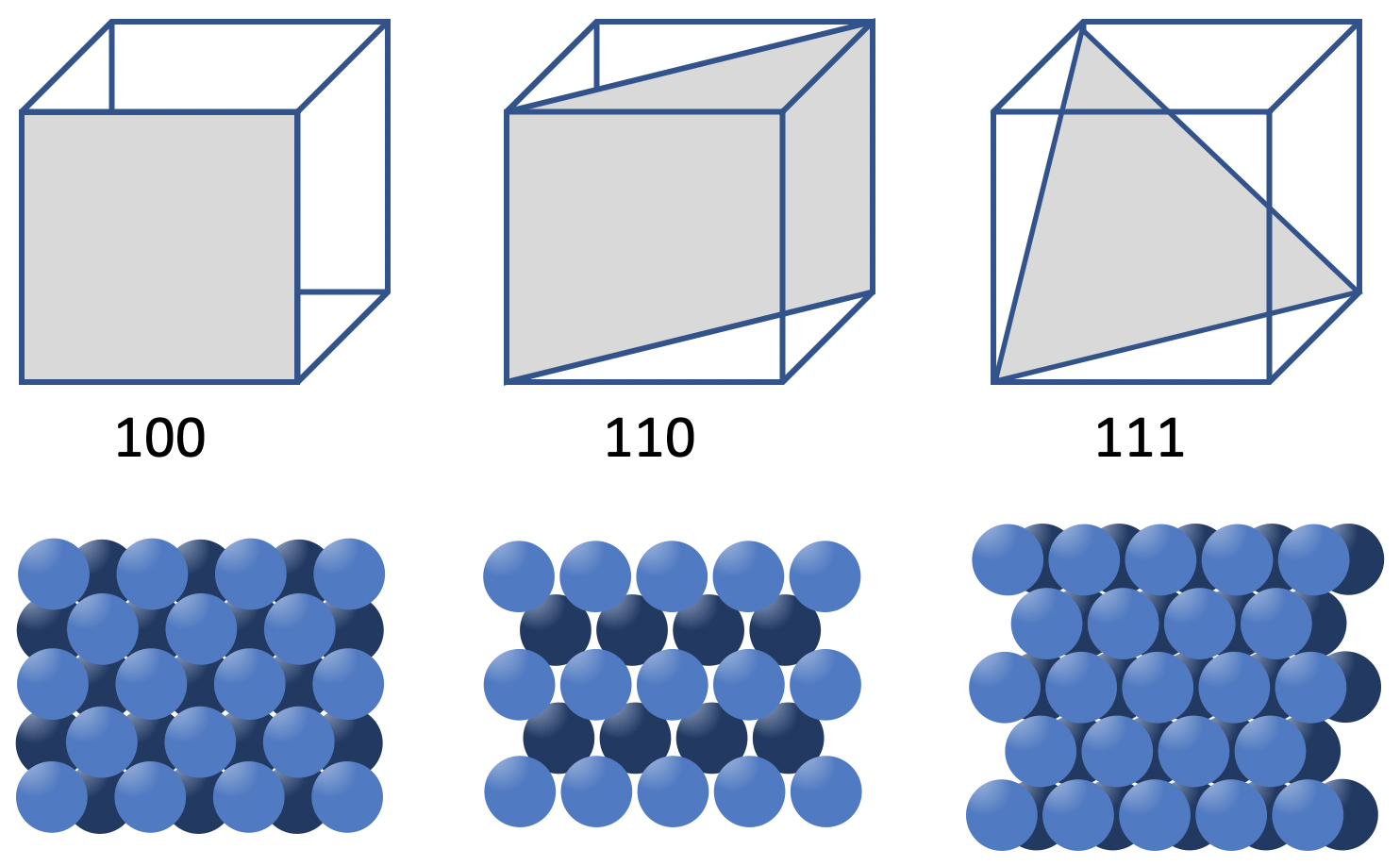}
	\end{center}
	\caption{Illustration of different surface structures (facets) using three different cutting orientations (100), (110) and (111). The three values (0 or 1) indicate the Miller indices for the cutting plane.}
	\label{fig:crystal}
\end{figure}

The binding site refers to selecting the atoms that bond between the adsorbate and the catalyst. To improve the likelihood of convergence for DFT we want to select initial positions that are as close as possible to the potential binding sites. A 2D example of bonding sites is shown in Figure \ref{fig:binding} between one, two and three atoms on the catalyst's surface and an adsorbate with two atoms. If the adsorbate has multiple atoms, the atom of the adsorbate that bonds must also be selected. We discuss the specific set of catalysts, adsorbates and binding sites used for relaxations in Section \ref{sec:datasets}.

\subsection{Classes of Materials for Electrocatalysts}
\label{sec:class_material}

There exists an enormous space of potential materials for electrocatalysts \cite{alonso2019fundamentals}. It would be overwhelming to catalog all the materials in one review. Instead, we provide some perspective on the space of materials to be considered. It is important to note that any materials dataset will necessarily contain limitations and bias.

Of the different types of materials, metal surfaces have received the greatest attention due to their reaction attributes, durability, and manufacture-ability. However, other materials are also possible, including (among many many others): surfaces with non-metallic elements, such as oxides (materials containing oxygen) \cite{seitz2016highly} or nitrides (materials containing nitrogen)\cite{abghoui2017onset}; metal-doped 2D materials such as graphene \cite{yang2016identification} or boronitrides\cite{grant2016selective};  Perovskites (metallic oxide minerals) \cite{hwang2017perovskites} that are more earth abundant and potentially less expensive; and other metal-free catalysts \cite{ma2019review}.  While these other material types show promise, we focus on metal surfaces in this paper given their potential for nearer term practical use and the clear applicability of DFT relaxations to their study. However, we hope the techniques developed for metal surfaces will also be useful for other material types. 

Figure \ref{fig:metalsurfaces} illustrates several different types of metal structures. These include pure metals, multi-metallics (multiple unordered metal alloys), intermetallics (ordered multi-metallics), 
single-atom catalysts in host materials \cite{giannakakis2018single}, and overlayers (surface layers of one or more metals). Alloys that contain more than three different metals are referred to as ``high-entropy alloys''. Historically, the most thoroughly studied are the simplest structures containing pure metals. In contrast, the space of multi-metallics, intermetallics and overlays is significantly less well explored given the vastness of the space. For instance, choosing up to three metals from the roughly 40 possible, results in almost 10,000 combinations, and this does not even include all the possible combinations of different ratios and patterns!

Given the same underlying bulk structure, there are multiple different surfaces that may be enumerated based on which orientation you slice the material. As shown in Figure \ref{fig:crystal}, the orientation of a crystalline surface or facet is typically indicated using coordinates on a unit cube. For instance, a diagonal facet is specified as (111) and results in the surface having a hexagonal pattern, whereas a (100) surface that cuts along a single dimension has a rectangular grid. In addition to flat surface types, surfaces may also contain steps, defects or other complications \cite{wang2019defect}. All of these variables impact the number of different bonding sites available and how many atoms may bond to the catalyst at one time (Figure \ref{fig:binding}), which in turn impacts the overall reaction rate.

\subsection{Other considerations}

While we covered many of the basics for modeling catalysts, numerous improvements could be made to further improve the realism of DFT calculations. While we do not cover these items in detail, we want to make the reader aware of these topics. For those interested in a more in-depth discussion, we recommend \cite{Norskov14}.

{\bf Coverage:} The coverage of a catalyst refers to the percentage of surface sites that are occupied by adsorbates. As the coverage increases, the adsorbates move closer together and impact the system's energy. For instance, the adsorption energy of the first oxygen placed on a Pt(111) surface is smaller than the adsorption energy of the 2nd, third, or fourth oxygen placed on the surface\cite{fiorin2009microcalorimetry}. Attractive and repulsive interactions between the adsorbates can also create ordered patterns of adsorbates on the surface \cite{mendez2005coadsorption}.

{\bf Poisoning:} Other species in the reacting mixture (even at trace levels) may poison the catalyst surface through a competitive adsorption process. Other adsorbates that do not participate directly in the reaction may adsorb strongly such that the coverage of the desired molecule and accompanying reaction rates are significantly reduced. Worse, they may bind so strongly that they never leave, thereby rendering the catalyst inert. These molecules occupy the potential bonding sites of the desired adsorbates, and essentially ``poison'' the catalyst. A common example is the poisoning of noble metal catalysts by lead, or poisoning of platinum catalysts by trace carbon monoxide.

{\bf Electrolytes:} Most electrochemical reactions occur in the presence of an aqueous or solid electrolyte to conduct charged particles to the surface, e.g. they are submerged in water/electrolyte solution to increase the conductivity of the system and therefore improve reactivity. The precise solvent (water or otherwise) can affect adsorption energies, most notably for adsorbates that can form hydrogen bonds with the water layer. Further, electrolytes such as KOH or NaCl can be added to the water to further improve reaction performance. They do this by interacting with the adsorbates and catalyst to stabilize the transition states of particular reaction intermediates, thereby reducing the activation energy and accelerating the reaction rate. Sometimes these effects are modeled by making simple corrections to the energies found without electrolytes, but ideally the electrolytes would be modeled directly. There are major scientific challenges remaining in our understanding of these processes. 

{\bf Catalyst Dynamics:} In the real world, the atoms in the system are constantly moving. The more heat that is applied to a system, the faster the atoms move. Molecular dynamics models the movement of these atoms by integrating the standard equations of motion ($F=ma$). Since the atoms may be fast moving and visiting a wider range of potential surfaces and sites, modeling these systems can be more challenging than the relaxations (with zero kinetic energy) we describe in this paper. The real catalyst may re-arrange under reaction conditions or form an ensemble of different states that all contribute to the final activity. Modeling these processes requires molecular dynamics simulations, but accessing experimentally-relevant timescales (nanoseconds) requires approximately 10$^6$ force calls at the atomic (femtosecond) timescale and are thus vastly more expensive than standard DFT relaxations. Approaches to exploring this problem use both standard DFT \cite{rapaport2004art} and learned models \cite{schutt2018schnet, li2015molecular,behler2016perspective,chmiela2017machine}.  

%% file: sections/ml.tex
\begin{figure*}
	\begin{center}
		\includegraphics[width=0.98\linewidth]{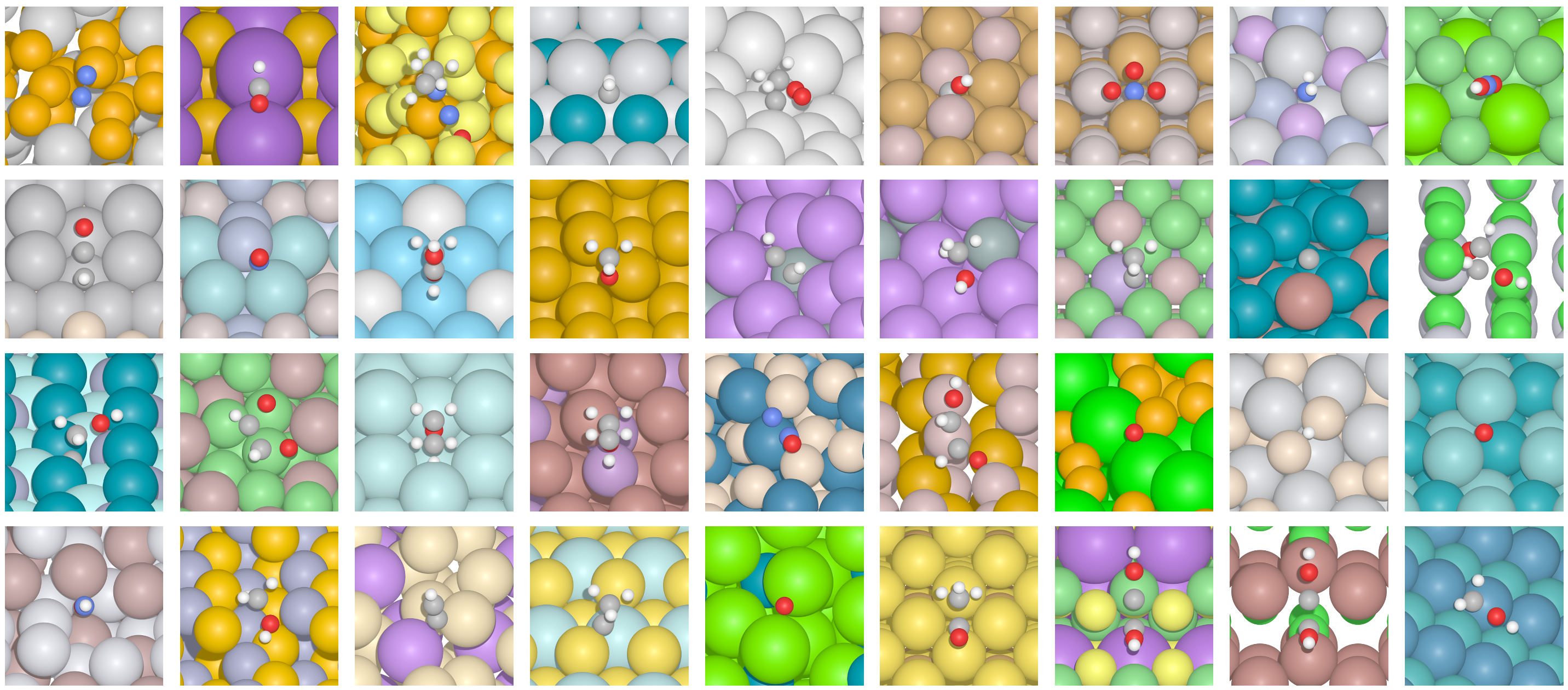}
	\end{center}
	\caption{Examples of cataylsts (larger atoms) and adsorbates (smaller atoms) in the \dataset{} dataset. Spheres represent the atoms with the color and size indicating the atomic number (element type) and the atomic radii, respectively. For example, the small white, red, grey, and blue spheres are hydrogen, oxygen, carbon and nitrogen, and the larger spheres are mostly metals.}
	\label{fig:examples}
\end{figure*}

\section{ML for Electrocatalysis}
\label{sec:ml}

In this section, we begin by providing a short overview of the problem, followed by the dataset, tasks, and approaches for addressing it. For those interested in a more detailed discussion of the problem, please refer to Sections \ref{sec:storage} and \ref{sec:electrocatalyst}.

As intermittent renewable energy sources (solar and wind) become more prevalent, the storage of their energy from times of peak generation to peak demand will increase in importance. The use of renewable energy to generate other fuels that are easily stored provides a potential solution to this problem that can scale to nation-sized grids. For instance, renewable energy can be used to split water into hydrogen and oxygen gas. The stored hydrogen is then used to generate electricity at a later time using fuel cells. Currently, a major limiting factor is finding efficient, durable and cheap catalysts for driving the electrochemical reactions involved in this process. 

A catalyst increases the rate of a chemical reaction while not being consumed in the process. During a chemical reaction the molecules involved in the reaction, called adsorbates, interact with the catalyst's surface. The change in energy when the adsorbate comes in contact with the catalyst, called the adsorption energy, is the key factor is determining the overall reaction's rate (Section \ref{sec:rates_in_practice}). The adsorption energy must be neither too high (the catalyst holds on to the adsorbates too strongly) nor too low (the adsorbates float away from the catalyst). If the reaction's rate is increased, the overall efficiency of the system increases and costs decrease. The challenge is finding catalysts that not only increase the reaction's rate, but are also cheap to produce, e.g., are not made of expensive metals such as platinum. 

The adsorption energy is computed by first finding the positions of the atoms when the adsorbate comes in contact with the catalyst's surface. In the real-world, the atoms seek a configuration that reduces the overall energy of the system. The set of 3D atom positions at a local energy minimum is referred to as a ``relaxed structure'' and the corresponding system energy as the ``relaxed energy''. 

How is a relaxed structure estimated computationally? An iterative process is used that first roughly places the adsorbate near the catalyst to create an initial structure. For any given pair of adsorbate and catalyst, numerous initial structures may exist based on the binding site chosen, Figure \ref{fig:binding}. The selection of an adsorbate's binding site and its initial positioning relative to it is heuristically determined. Next, the per-atom forces are computed, from which atom positions are updated using BFGS~\cite{head1985broyden} or conjugate gradient techniques. The forces are re-estimated and atom positions updated until the forces converge to zero at which point a local energy minimum has been found. Typically, hundreds of iterations are necessary to converge to a relaxed structure. While this process seems simple, the computation of the atom forces is highly complex and computationally very expensive. 

Atom forces and a structure's energy are computed using Density Functional Theory (DFT) \cite{sholl2011density, parr1980density}, which is computationally expensive and can take hours to calculate for a single structure. How can we scale the simulation of new catalysts from hundreds or thousands to billions? A promising alternative to expensive DFT calculations is to train efficient machine learning models to approximate DFT. The model takes as input the set of atoms, including their atomic numbers and 3D positions, and computes the per-atom forces. Similar to DFT, the force estimates may be used to find relaxed structures. To eliminate the need for DFT entirely, models may be trained to also estimate a structure's energy. 

Machine learning models that employ deep learning require large datasets for training. We begin by describing the Open Catalyst Dataset (\dataset{}) created for this purpose; followed by deep learning approaches that have been proposed for estimating DFT and metrics for evaluating the models. We strive to release all data and code in accordance with FAIR (Findable, Accessible, Interoperable, Reusable) principles \cite{wilkinson2016fair}.

\begin{figure*}
	\begin{center}
		\includegraphics[width=0.98\linewidth]{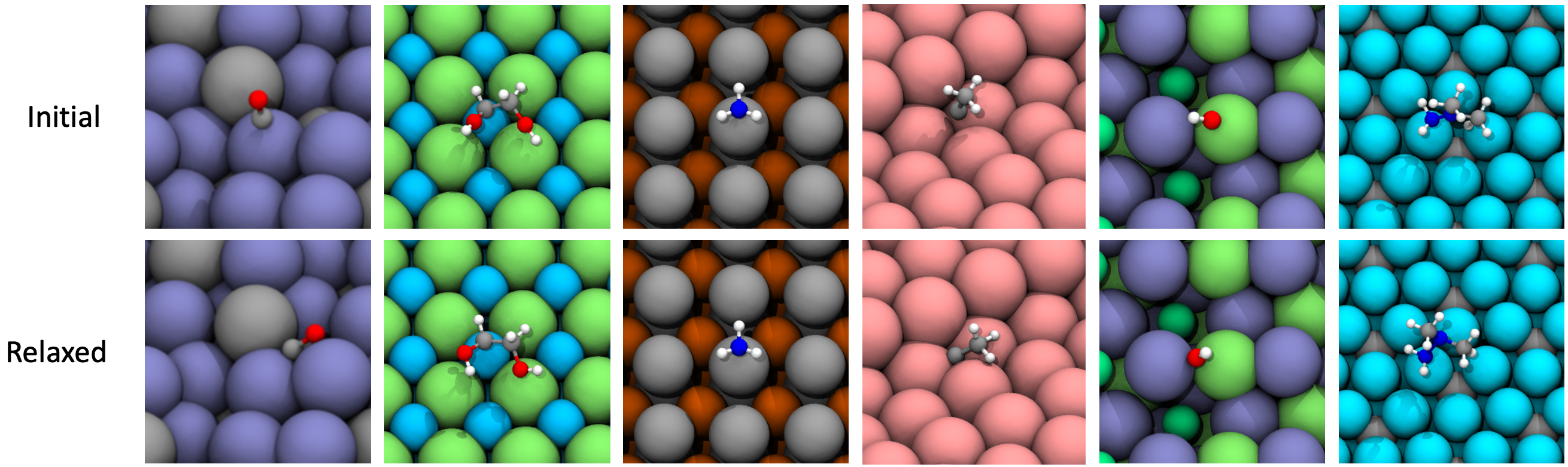}
	\end{center}
	\caption{Example initial and relaxed structures from the \dataset{} dataset. Note that the changes from initial to relaxed structures can appear to be quite small, especially for adsorbates containing only a few atoms. However, the changes in energy can be significant.}
	\label{fig:relaxation}
\end{figure*}

\subsection{Open Catalyst Dataset (\dataset{})}
\label{sec:datasets}

An ideal dataset would enable the exploration of new techniques for estimating DFT calculations while also helping the catalysis community explore new materials and gain a deeper understanding of the physics underlying catalysts---overall accelerating progress towards discovering better alternatives to commonly used catalysts. This requires both a large (for learning) and diverse (for generalization and exploring new materials) dataset. For the dataset, priority is given to catalysts with relatively near-term practical potential, and adsorbates relevant to the reactions used in renewable energy storage. Specifically, we explore metallic surfaces, which are currently the most popular catalysts in use due to their durability, ease of manufacture, and efficiency. 

Our \dataset{} dataset \cite{OCD1M20} contains over a million relaxations of randomly selected catalysts and adsorbates from a set of potential candidates. On average each relaxation contains approximately 200 structures in its trajectory from the initial structure to the relaxed structure, which results in over 130+ million structures for training with the rest used for validation and testing. Example adsorbate and catalyst pairs are shown in Figure\ref{fig:examples}, and example initial and relaxed structures in Figure \ref{fig:relaxation}. All calculations are performed using VASP (Vienna Ab initio Simulation Package) \cite{kresse1993ab,kresse1994ab,kresse1996efficiency,kresse1996efficient}, which is a popular package for performing the types of DFT calculations used in this paper.

A detailed description of the \dataset{} dataset and its various design choices are provided in \cite{OCD1M20}. In this paper, we cover the necessary details to build and evaluate models with the dataset. We begin by describing the various tasks, followed by the inputs and data splits.

\begin{figure}
    \centering
    \includegraphics[width=\columnwidth]{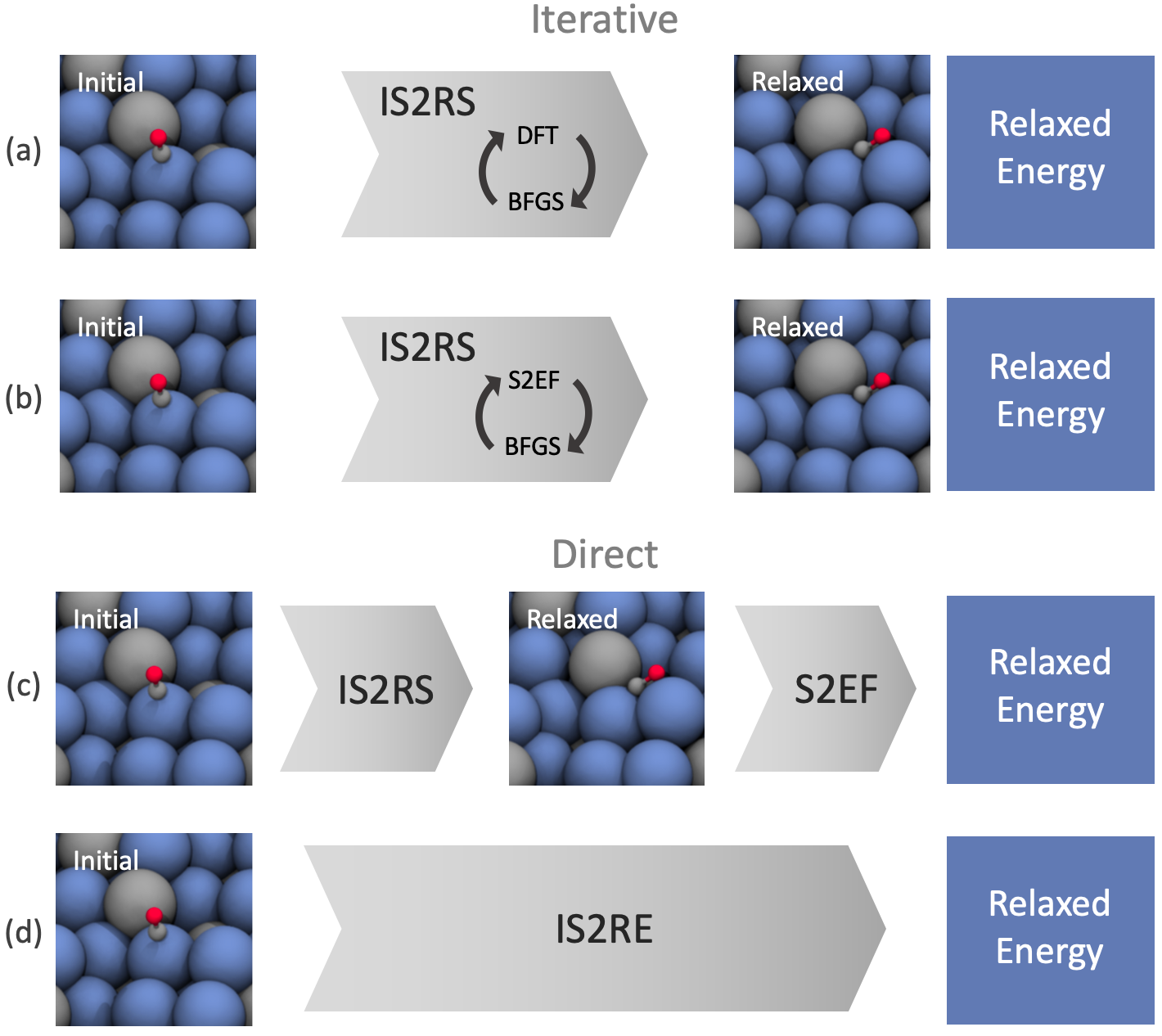}
    \caption{Illustration of the relationship between tasks in computing a relaxed structure or relaxed energy from an initial structure. (a) the traditional approach uses DFT along with a linear optimizer, such as BFGS or conjugate gradient, to iteratively update the atom positions until the relaxed structure and energy are found. (b) using ML models trained to predict the energy and forces of a structure, S2EF can be used as a direct replacement for DFT. (c) the relaxed structure could potentially be directly regressed from the initial structure and S2EF used to find the energy. Finally, if only the relaxed energy is desired, (d) could compute the relaxed energy directly from the initial state.}
    \label{fig:tasks}
\end{figure}

\subsubsection{Tasks and Evaluation Metrics}
\label{sec:tasks}
We identify three tasks based on their practical applicability to cataylst design. Since structure relaxations are the primary tool for determining a catalyst's utility, we focus on modeling this process and its outputs. As shown in Table \ref{tab:IO}, task outputs include per atom forces and positions, and the energy of a structure. The relationship of each of our tasks is shown in Figure \ref{fig:tasks}. Per task evaluations may be performed on the validation set using the supplied code and ground truth data. Test set evaluations may be performed by uploading results to the \href{http://www.opencatalystproject.org}{Open Catalyst Project leaderboard}. Our three tasks are as follows:

\paragraph{S2EF - Structure to Energy and Forces} task is given a structure as input, predict the per-atom forces and structure's energy as output. This task mirrors the computationally expensive DFT calculation. An efficient approximation to DFT would be beneficial to numerous tasks, including relaxations. Evaluation is performed using the Mean Average Error (MAE) between the computed forces and energies and those computed using DFT. In addition, a third evaluation metric determines the percentage of predictions with Energy and Forces within a Threshold (EFwT) of the ground truth. By using tight thresholds, the EFwT metric provides an estimate for how often the predictions would be practically useful. 

\paragraph{IS2RS - Initial Structure to Relaxed Structure} task estimates the relaxed structure given the initial structure as input. This task may be performed iteratively using S2EF calculations, similar to how traditional relaxations are computed, Figure \ref{fig:tasks}(b). We refer to these calculations as ML relaxations, since ML models are used to compute the relaxations. An alternative is to compute the final 3D positions of the atoms directly from their initial structure, Figure \ref{fig:tasks}(c).

Evaluations are performed using two techniques. The first measures the 3D positional differences between the ML computed relaxed structure and the structure computed using a traditional DFT relaxation. Since small changes in position can lead to significant changes in forces and energies, distance metrics by themselves are insufficient for determining whether a true relaxed structure is found. The second metric uses a single DFT calculation to determine the forces on the proposed relaxed structure. If a true local minimum was found, the forces as computed by DFT should be close to zero. The final metrics compute the percentage of ML relaxed structures with maximum force magnitudes below a threshold. Since DFT calculations are computationally expensive, IS2RS force metrics reported on the Open Catalyst Project test server are only evaluated on a subset of the test splits. The 3D distance metric is computed using the entire test set.

\begin{figure*}
	\begin{center}
		\includegraphics[width=0.98\linewidth]{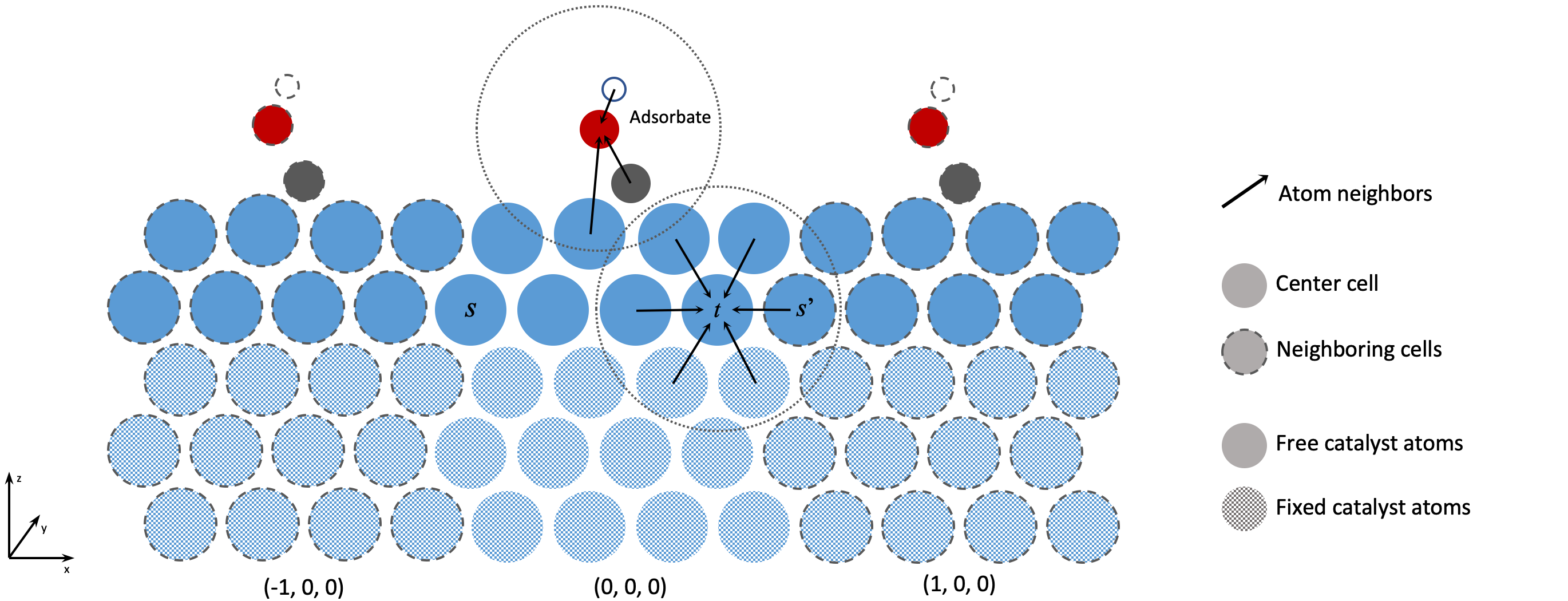}
	\end{center}
	\caption{2D Illustration of a slab that represents a catalyst's surface and an adsorbate. The slab is tiled in the X and Y directions to create the surface (neighboring cells shown as atoms with dashed outlines). Only the cells to the left ($[-1, 0, 0]$) and right ($[1, 0, 0]$) are shown. The adsorbate is also assumed to be tiled with the slab (white, red, and grey atoms). Only the top 2 layers of the slab are allowed to move during a relaxation (dark blue), and the others are fixed (light blue). Neighboring atoms (black arrows) can be from the same cell or neighboring cells ($t$ and $s'$). All atoms within a radius (dotted circle) are assumed to be neighbors.}
	\label{fig:slab}
\end{figure*}

\paragraph{IS2RE - Initial Structure to Relaxed Energy} task computes the relaxed structure's energy from the initial structure. The relaxed energy is used to compute the adsorption energy, i.e., the change in energy when the adsorbate comes in contact with the catalyst. As discussed in Section \ref{sec:rates_in_practice}, the adsorption energy is informative in predicting the activation energies and ultimately the rate of the reaction in the presence of the catalyst. While we focus on the system's energy, other system properties may also be estimated. For instance, the popular QM9 dataset \cite{ramakrishnan2014quantum} containing small organic molecules, estimates 13 energetic, electronic, and thermodynamic properties.

Similar to IS2RS, IS2RE may be accomplished using two approaches: The first iteratively estimates the relaxed structure and its corresponding energy using S2EF, Figure \ref{fig:tasks}(b). The second, directly computes the relaxed energy from the initial structure, Figure \ref{fig:tasks}(d). Using the second approach, IS2RE does not require the estimation of the relaxed structure. Evaluation is performed using MAE between the estimated relaxed energies and those found using DFT relaxations, as well as, the percentage of energy estimates within a tight threshold of the ground truth.

\subsubsection{Inputs}
The inputs to the \dataset{} tasks include the atoms, catalyst and adsorbate information. We discuss each in turn.

\paragraph{Atoms} A structure contains catalyst and adsorbate atom information at a specific instance in time. This includes the atoms' 3D positions, along with various per-atom and per-pair (of atoms) proprieties, as shown in Table \ref{tab:IO}. Per-atom information includes the atom's group and period number, electronegativity, covalent radius, etc. The per-pair properties include the distance between the atoms and their predicted bond type.

\paragraph{Catalyst} Along with the atom information described above, additional information is provided for the catalyst. Each catalyst contains 1 to 3 types of elements, with $5\%$, $65\%$, and $30\%$ of the \dataset{} catalysts being unary, binary, and ternary materials respectively. Stable catalyst materials were randomly selected from those contained in the Materials Project~\cite{jain2013commentary}. A catalyst is modeled as an infinite 3D surface with a repeating 2D crystalline structure in the x and y dimensions. The adsorbate sits on top of the catalyst in the z dimension with the catalyst extending infinitely below the adsorbate. As input, only a single crystalline structure, known as a \gls{slab}, is given that can be replicated or tiled in cells to create the full 3D surface. A 2D illustration of this process is shown in Figure \ref{fig:slab}. When computing atom interactions, the set of neighboring atoms needs to be determined. Since the slab is repeated in each cell, the atoms on the far right of a slab (atom $t$ in Fig.~\ref{fig:slab}) may be neighbors to those on the far right of the next cell over (atom $s'$ and $s$ in Fig.~\ref{fig:slab}), and similarly for all other x and y directions. The atom positions can be transformed from one cell to the next using a $3\times3$ matrix $\mathbf{M}$:
\begin{equation}
    \mathbf{x}^* = \mathbf{M}\mathbf{c} + \mathbf{x},
\end{equation}
where $\mathbf{c}$ is the 3D cell indicator vector, e.g., $[1, 0, 0]^{\mathbf{T}}$ or $[-1, 1, 0]^{\mathbf{T}}$, $\mathbf{x}$ is the atom's position in the slab, and $\mathbf{x}^*$ is the position of the atom in the indicated cell. 

Instead of tiling the slab in the z direction, it is assumed that only atoms near the surface of the catalyst will move. Each atom in the structure is given a variable that indicates whether it is assumed to be fixed or is free to move. All atoms in the adsorbate are free, and typically the top two layers of atoms in the catalyst are free and the rest fixed (dark and light blue atoms in Fig. \ref{fig:slab}). Note that DFT computes the forces assuming that the slab is not tiled in the z direction, so the forces of the bottom atom layers in the catalyst may be very large as they try to expand. Only free atoms are used when evaluating computed positions and forces. The forces of the fixed atoms may be used for training if desired. 

\begin{table*}[t]
\centering
\begin{tabular}{ |>{\centering\arraybackslash}m{6cm}|m{4cm}|m{6cm}| }
 \hline
 Reaction & Name & Purpose \\ \hline \hline
 \ce{2H+ + 2e- -> H2} & Hydrogen Evolution Reaction (HER) & Conversion of renewable energy to H$_2$/O$_2$\\ \hline
 \ce{2H_2O ->  O2 + 4H+ + 4e- } & Oxygen Evolution Reaction (OER) & Conversion of renewable energy to  H$_2$/O$_2$\\\hline
   \ce{H_2 -> 2H+ + 2e- } & Hydrogen Oxidation Reaction (HOR) & Conversion of hydrogen to electricity\\\hline
   \ce{4e^- + O2 + 4H+ -> 2H_2O  } & Oxygen Reduction Reaction (ORR) & Conversion of hydrogen to electricity\\\hline
 \ce{CO2 + 4H2 +  -> CH4 + 2H2O} & Methanation & Production of methane (main component of natural gas) and \ce{CO2} reduction \\ \hline
  \ce{N2 + 3H2 -> 2NH3} & Haber-Bosch & Production of ammonia for fertilizer \\ \hline
    \ce{N2 + 6H+ + 6e- -> 2NH3} & Nitrogen Reduction Reaction (NRR) & Electrochemical production of ammonia for fertilizer from renewable energy\\ \hline
    \ce{ 2NH3 -> N2 + 3H2} & Ammonia Decomposition & Ammonia as a hydrogen carrier \\ \hline
  \ce{CO2 + ne- + nH+ -> \text{products}} & CO$_2$ Reduction Reaction (CO$_2$RR)  & Renewable chemical building blocks from CO$_2$  \\ \hline
  \ce{CO + H_2 -> \text{products} } & Fischer-Tropsch & Production hydrocarbons from renewable H$_2$/CO\\ \hline
  \ce{CO + H_2O -> CO2 + H2} & Water Gas Shift (WGS) & H$_2$ production from CO$_2$ \\ \hline
    \ce{C2H6 -> C2H4 + H2} & Non-Oxidative Ethane Dehydrogenation & Upgrading of waste ethane \\ \hline


\end{tabular}
\caption{Subset of the many reactions that are relevant to the \dataset{} dataset.}
\label{tab:reactions}
\end{table*}

\begin{figure}
    \centering
    \includegraphics[width=\columnwidth]{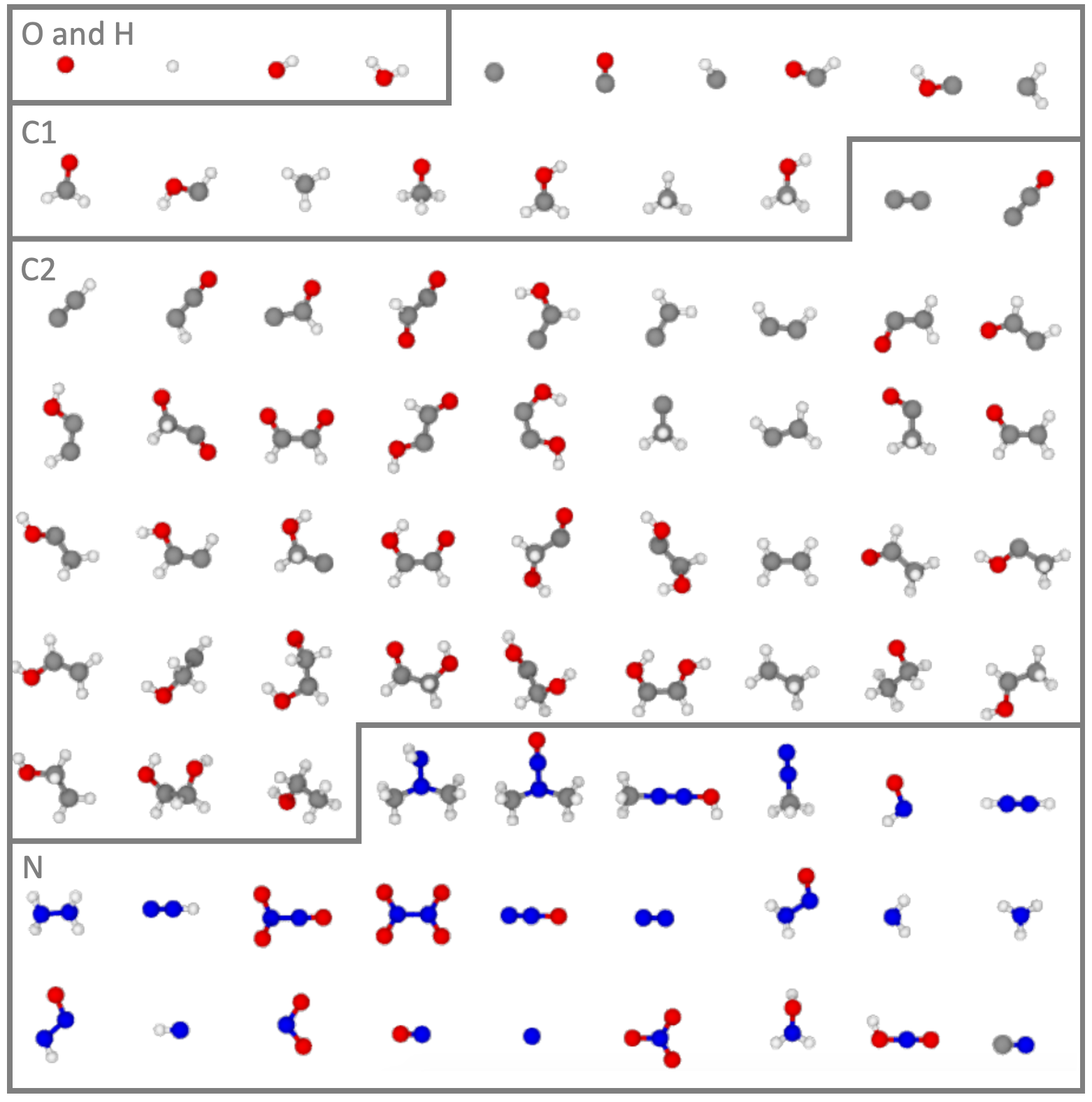}
    \caption{The adsorbates used in the \dataset{} dataset \cite{OCD1M20} contain oxygen (red), hydrogen (white), one carbon (grey), two carbon (gray) and nitrogen (blue) molecules useful for renewable energy applications.}
    \label{fig:adsorbates}
\end{figure}

\paragraph{Adsorbate} The adsorbates are randomly chosen from the 82 molecules shown in Figure \ref{fig:adsorbates}. The set of adsorbates was selected based on the molecules, called intermediates, that are involved in the reactions of interest, such as those in Table \ref{tab:reactions} for renewable energy storage and sustainable agriculture. The adsorbates are clustered into four groups:  small molecules containing a few atoms, small molecules with at least one carbon (C\textsubscript{1}), larger molecules with at least two carbons (C\textsubscript{2}), and molecules containing nitrogen. The study of these adsorbates has significant practical value. For instance, analysis of C\textsubscript{1}/C\textsubscript{2} intermediates will allow new insights into a catalyst's selectivity (do they produce any undesired byproducts) for CO\textsubscript{2} conversion to fuels. Note there is a larger number of high-complexity (i.e., larger) adsorbates. By sampling from the adsorbates uniformly, we implicitly perform more calculations on the larger adsorbates for which there is less prior data. This helps the catalysis research field by generating more previously unexplored data, and provides a more diverse training dataset for the ML community.

\begin{table}[t]
\centering
\begin{tabular}{ |ll| }
    
        \multicolumn{2}{c}{Inputs} \\
    \hline
        \multicolumn{2}{|c|}{Per atom} \\ \hline
        atom positions & 3D $(x, y, z)$ positions   \\
        group number & column in the periodic table ($1$-$18$) \\
        period number & row in the periodic table ($1$-$9$) \\
        electronegativity & tendency to attract electrons \\
            & ($0.5$-$4.0$) \\
        covalent radius~\cite{cordero2008covalent} & size of an atom part of one \\
            & covalent bond ($25$-$250$ pm) \\
        valence electrons & $\#$ of outer shell electrons ($1$-$12$) \\
        first ionization energy & energy needed to remove \\
            & outermost electron ($1.3$-$3.3$ eV) \\
        electron affinity & energy released when electron \\
            & attaches to a neutral atom \\
            & to form a negative ion ($-3$-$3.7$ eV) \\
        block & s, p, d, f \\
        atomic volume & $1.5$-$4.3$ cm$^3$/mol \\
    \hline
        \multicolumn{2}{|c|}{Per pair} \\ \hline
        bond type & single, double, etc. (one-hot)\\ 
        distance & 3D $(x, y, z)$ delta positions\\ \hline
    \noalign{\medskip}  
        \multicolumn{2}{c}{Outputs} \\
    \hline
        \multicolumn{2}{|c|}{Per atom} \\ \hline
        atom positions & 3D $(x, y, z)$ positions   \\
        atom forces & 3D $(f_x, f_y, f_z)$ values   \\
    \hline
        \multicolumn{2}{|c|}{Per system} \\ \hline
        energy & $E$ energy \\ \hline
\end{tabular}
\caption{Summary of model inputs and outputs.}
\label{tab:IO}
\end{table}

\subsubsection{Train, Validation, and Test Splits}

The \dataset{} dataset contains over 1.3 million relaxations that required over 70 million hours of compute \cite{OCD1M20}. The data is split into train, validation and test sets. Each relaxation contains the 3D atom positions, atom forces, and energies for the initial, intermediate, and final relaxed structures. The training dataset contains over 650,000 relaxations with over 130 million structures. 

A model's ability to generalize may be evaluated using different validation and test splits. Generalization to new catalysts and adsorbates is one of high practical impact, since this would enable the application of models to new catalyst materials and chemical reactions. We propose creating validation and test splits specifically for these scenarios. We create four subsplits for validation and testing based on whether the catalyst material or adsorbate is present in the training dataset. A catalyst material is said to be in the training dataset if a catalyst containing the same elements is contained in the training dataset, even if the elements are in different arrangements or proportions. The four subsplits are labeled In Domain (same distribution as training dataset), Out of Domain Adsorbate (OOD Adsorbate), OOD Catalyst and ODD Both (neither the adsorbate or catalyst material is in training). Each subsplit contains approximately 70,000 to 90,000 relaxations. For the S2EF task we randomly select a one million structure subset from the relaxations in each subsplit. A random subset of 25,000 relaxations per split is sampled for the IS2RS and IS2RE tasks. The subsplits of our validation set are exclusive to the subsplits in the test set, e.g., the adsorbates in the validation OOD Adsorbate subsplit are unique from the adsorbates in the test OOD Adsorbate subsplit and similarly for OOD Catalyst and OOD Both.

The train and validation datasets are provided to the public. The inputs to the test dataset are also provided. However, the test outputs are not made public to ensure overfitting to the test set does not occur. To evaluate on the test dataset, results can be uploaded to the \href{http://www.opencatalystproject.org}{Open Catalyst Project leaderboard}.

\subsection{Models}
\label{sec:models}

Our goal is to help enable the community to develop models that can predict a structure's energy and per atom forces (S2EF), relaxed structure (IS2RS) and relaxed energy (IS2RE). High practical impact could be realized if efficient solutions to these problems can be found. Currently, several approaches have been proposed for addressing our three tasks. The results of several state-of-the-art approaches on \dataset{} can be found in \cite{OCD1M20} or on the \href{http://www.opencatalystproject.org}{public leaderboard}. 

Numerous deep learning approaches have been proposed for predicting the properties of a structure \cite{behler2007generalized,duvenaud2015convolutional,schutt2017schnet,gilmer2017neural, jorgensen2018neural, xie2018crystal, qiao2020orbnet, klicpera2020directional, nachmani2020molecule}. One challenging problem is how to model the interactions of the atoms. Since the atoms do not lie on a regular grid, such as pixels in an image, standard approaches such a convolutional neural networks \cite{lecun1998gradient} cannot be applied. Instead, Graph Neural Networks (GNNs) \cite{gori2005new,li2015gated,zhou2018graph} are commonly used \cite{schutt2017quantum,gilmer2017neural,jorgensen2018neural,schutt2017schnet,schutt2018schnet,xie2018crystal,qiao2020orbnet}. A graph is represented by a set of vertices and edges where each vertex represents an atom and the edges represent the interactions between the atoms. Computation on the graph is done by updating each vertex's (atom's) $x$'s hidden representation $h_x$. The updates iteratively pass messages between vertices connected by the edges:
\begin{equation}
    h_x^{t+1} = g(h^{t}_{x \cup N(x)}, v_{x \cup N(x)}, e_{x \leftrightarrow N(x)})
\end{equation}
$N(x)$ is the set of neighboring atoms of $x$. The initial states $v$ of the vertices are based on the atom features, such as element type or atomic numbers, and the edge features $e$ are computed from the distances between atoms and potentially other bond features. $v_{x \cup N(x)}$ refers to the initial states of atom $x$ and its neighbors $N(x)$, and $e_{x \leftrightarrow N(x)}$ refers to features of edges with $x$ as an endpoint.

The structure property $y$ of interest, such as the energy, is typically estimated locally at each vertex and averaged or summed together for the final result:
\begin{equation}
y = \sum_x g'(h_x)
\end{equation}
Neural networks are trained for functions $g$ and $g'$ using a regression loss function. An overview of early message passing techniques and how they relate can be found in \cite{gilmer2017neural}. To the best of our knowledge, the first to apply a variant of GNNs to materials was SchNet \cite{schutt2017schnet}.

Improvements to these models include new methods for adding edges, such as using Voronoi tessellation or K-nearest neighbors \cite{jorgensen2018neural,xie2018crystal}, and increasing the amount of information used in message passing \cite{jorgensen2018neural}. The use of continuous instead of discrete filters \cite{schutt2017quantum,gilmer2017neural} for pairwise interactions was proposed by \cite{schutt2017schnet,schutt2018schnet} that enabled their use in molecular dynamics. An exploration of the tradeoffs between equivariant and invariant pairwise information is provided by \cite{miller2020relevance}. As noted by \cite{anderson2019cormorant}, networks should be rotationally invariant and they propose explicitly enforcing that features are covariant to rotations. In \cite{chen2019path,velivckovic2017graph}, global attention layers are used to model higher-order graph properties, while \cite{klicpera2020directional} use directional message passing to model angular information between triplets of atoms. The limits of modeling higher-order terms by only looking at 3 or 4 body correlations is explored in \cite{PhysRevLett.125.166001}.  A promising approach to achieving higher accuracy while improving efficiency is the incorporation of more domain specific features, such as those obtained from mean-field electronic structure methods \cite{qiao2020orbnet}, or atomic orbital interaction features for crystal structures \cite{PhysRevMaterials.4.093801}. 

The per-atom forces for the S2EF task have been computed using two approaches. The first is to estimate directly the forces $f_x \in \mathbb{R}^3$ for atom $x$ in the same manner as the system properties as described above \cite{li2015molecular}:
\begin{equation}
f_x = g(h_x)
\end{equation}
However, estimating each atom's forces in isolation may lead to inconsistencies between energies and forces, i.e., forces along a closed path may not sum to zero \cite{chmiela2017machine}.

The second approach is to estimate the forces directly from the energy by taking the
partial derivative of the energy $E$ relative to the atom's positions $p_x$
\cite{schutt2017schnet,schutt2018schnet}:
\begin{equation}
    f_x = -\frac{\partial E}{\partial p_x}(v, e)
\end{equation}
A loss based on the computed forces may also be added during training \cite{schutt2018schnet}. This approach makes the assumption that the network is differentiable with respect to the atom positions. For instance, the pairwise features that encode the distances between atoms should not use a discrete (bucket) representation \cite{schutt2017quantum} that results in discontinuities in the derivatives. Instead continuous distance filters are a better choice \cite{schutt2017schnet,schutt2018schnet}.

Research in predicting the relaxed structure from the initial structure (IS2RS) through the use of ML models is still an emerging area of research. Please see the \href{http://www.opencatalystproject.org}{\dataset{} public leaderboard} to view evaluations on the latest models. If relaxed structures are estimated through an ML relaxation process (Figure \ref{fig:tasks}(b)), the forces estimated  (S2EF) at each iteration need to be of high accuracy. While the accuracy of energy and force predictions are improving, as shown in \cite{OCD1M20}, less than $1\%$ of S2EF structure calculations performed by baseline models (SchNet \cite{schutt2018schnet}, CGCNN \cite{xie2018crystal}, DimeNet \cite{klicpera2020directional}) are within accuracy thresholds used to estimate practical utility. Not surprisingly, when these force predictions are used in the IS2RS task, less than $1\%$ of relaxed structures are similarly found to be practically useful. While significant room for improvement exists, it is encouraging that lower force prediction errors does lead to improvements in the prediction of relaxed structures\cite{OCD1M20}. Furthermore, while ML relaxations may not at first accurately estimate relaxed structures, they may still be useful in speeding up DFT relaxations. For instance, the output of an ML relaxation may be used as the initial structure for a DFT relaxation, for which significantly fewer iterations (DFT calculations) may be needed to converge. 

Another approach to the IS2RS task is to regress the atom positions directly from an initial structure. While this may still be worth testing, it is unlikely to be effective. In typical regression problems, a neural network will regress to the mean. For the problem we are addressing, the mean of several local minima is unlikely to be a minimum itself (the mean of an atom sitting on either side of another atom will place both atoms in the same location).

For the IS2RE task the same approach as S2EF is typically used, except the initial structure is given as input. However, the accuracy of the estimated relaxed energy has been shown to be significantly worse if the initial structure is given as input rather than the relaxed structure \cite{gilmer2017neural}.  Thus, one hypothesis is that the estimation of the relaxed structure is necessary to determine the relaxed energy accurately.

\subsection{Related Datasets}

Datasets for catalysis are growing in popularity, especially as the use of machine learning models has increased. These datasets range in size from a few thousand to ~90,000 reaction and adsorption energies. An online collection of over 50 datasets can be found on Catalysis Hub \cite{winther2019catalysis}, containing approximately 109,000 computed energies. Included in this collection is a dataset \cite{mamun2019high} that computed ~90,000 relaxations to find adsorption and reaction energies for 2,035 binary alloy surfaces with 12 adsorbates. Datasets for intermetallic surfaces with small molecule adsorbates include those released by \cite{tran2018active} and  \cite{tran2019methods} that contain 40,000 and 47,000 energies respectively using approximately 2,000 catalyst surfaces. Datasets containing less than a few thousand calculations include those for (111) transition metal surfaces \cite{schumann2018selectivity} and high-entropy alloys \cite{batchelor2019high}, while others focus on specific reactions such as the \ce{CO2} reduction reaction \cite{ma2015machine}. If considering bulk material properties (not surfaces or adsorbates), \cite{dunn2020benchmarking} is a benchmark across 13 diverse material properties with over 100,000 samples, OQMD~\cite{saal2013materials} contains information on over 600,000 materials, and AFLOW~\cite{curtarolo2012aflow} studies millions of materials.

Information covering a broad set of materials and molecules can be found in the Materials Project \cite{jain2013commentary}. The catalyst materials for \dataset{} are sampled from those in the Materials Project.

%% file: sections/discussion.tex
\begin{figure*}
	\begin{center}
		\includegraphics[width=0.8\linewidth]{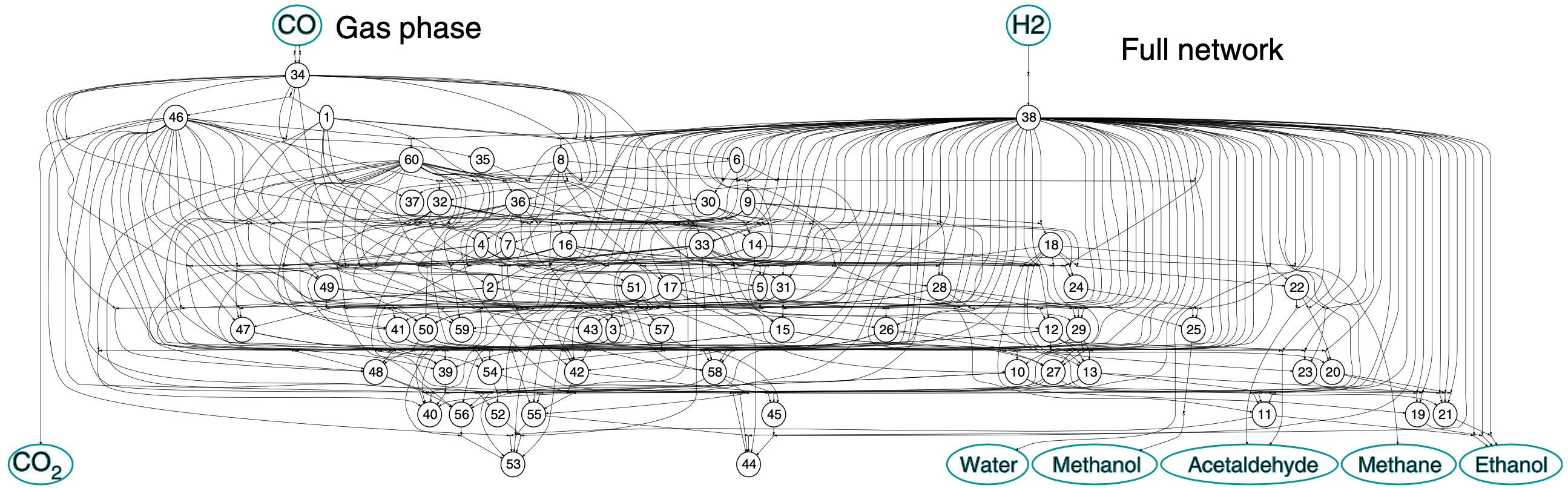}
	\end{center}
	\caption{Diagram of the over 2,000 possible reaction pathways from syngas (\ce{CO}+\ce{H2}) to \ce{CO2}, water, methane, methanol, acetaldehyde and ethanol. Determining a catalyst's impact across numerous reaction pathways remains an open question. Figure adapted from \cite{ulissi2017address}.}
	\label{fig:pathway}
\end{figure*}

\section{Discussion}
\label{sec:discussion}

Training efficient and accurate models for approximating DFT is the first step in using machine learning to aid in the design of new electrocatalysts. If this is accomplished, numerous potential areas for future research exist. Currently, the understanding of how larger molecules interact with an catalyst is limited. The production of fuels with higher energy densities depends on modeling these larger (C2 and C3) molecules \cite{li2018short}. In these molecules, it is likely that multiple atoms (the carbons) bond with the surface of the catalyst, creating a much larger and complex search space when computing potential energies. Efficient and accurate approximations could lead to to new insights for these reactions.

For many reactions, numerous reaction pathways exists. The large set of possible reaction pathways for the relatively simple reaction of syngas (\ce{CO}+\ce{H2}) is shown in Figure \ref{fig:pathway} \cite{ulissi2017address}. The pathways realized in practice, which impacts the level of selectivity (percentage of desirable products), are dependent on the choice of catalyst. For example, a catalyst might increase a pathway's rate that produces methane. However, it may also result in increasing in the rate of another pathway that produces the unwanted byproduct \ce{CO2}. This decrease in the desired selectivity of the catalyst reduces its practical usefulness. The design of new catalysts that optimize their impact on numerous reaction pathways is an open and challenging problem.  

Developing more realistic simulations is of high importance to the catalyst community. These simulations typically require modeling a larger number of atoms than is currently feasible with DFT. The types of interactions that could be modeled include:  the interactions of adsorbates with each other, the use of electrolytes \cite{wang2018electrolyte} to increase electron transfer, promoters \cite{vayenas2001electrochemical} to increase the reaction rate, and the poisoning of catalysts \cite{stonehart1972effect}. It remains an open question whether ML approximations of DFT can scale to accurately handle larger systems.  

In summary, efficient and accurate approximations to DFT could lead to the discovery of new catalysts for use in the cost-effective storage of renewable energy and reducing climate change. We explored different methods for energy storage and discussed their scalibility to nation-sized grids. We introduced the basic concepts behind the design of catalysts and how DFT is utilized in this process. Finally, we described how the \dataset{} dataset may be used to train ML models for approximating DFT. We are hopeful that research at the intersection of AI and catalysis offers both interesting scientific challenges and the opportunity for long-term practical impact on pressing global problems.